\documentclass{aa}  
\usepackage[colorlinks=true, allcolors=blue]{hyperref}
\usepackage{array}
\usepackage{multirow}
\usepackage{graphicx}
\usepackage{subfigure}
\usepackage{float}
\usepackage{figsize}
\usepackage{txfonts}
\usepackage{booktabs}
\usepackage{natbib}
\usepackage{siunitx}
\usepackage{amsmath}
\usepackage{relsize} 

\usepackage{subfigure}
\usepackage{comment}
\usepackage{natbib}

\def\aap{{\it A\&A}}

\def\apjl{{\it ApJL}}

\newcommand\simlt{\lower.5ex\hbox{$\; \buildrel < \over \sim \;$}}
\newcommand\simgt{\lower.5ex\hbox{$\; \buildrel > \over \sim \;$}}

\begin{document} 
\title{\textcolor{blue}{Evaluating Quasi-Periodic Variations in the $\gamma$-ray Lightcurves of Fermi-LAT Blazars}}
\titlerunning{Evaluating QPOS....}
   \author{F. Ait Benkhali
          \inst{1},
          W. Hofmann
 	        \inst{1},
          F.M. Rieger
           \inst{1,2}
           \and
          N. Chakraborty\inst{1,3}	
         	}
        
  \institute{ $^1$ Max-Planck-Institut f\"ur Kernphysik, Saupfercheckweg 1, 69117 Heidelberg, Germany \\
   				$^2$ ZAH, Institut f\"ur Theoretische Astrophysik, Universit\"at Heidelberg, Philosophenweg 12, 
                     69120 Heidelberg, Germany \\
              $^3$ Data Assimilation Research Centre, University of Reading, Whiteknights Rd, Reading RG6 6BB, United Kingdom        
                     }

  \abstract
   { 
The detection of periodicities in the light curves of active galactic nuclei (AGN) could have profound consequences for our understanding of the nature and radiation physics of these objects. 
At high energies (HE; $E>100$ MeV) five blazars (PG 1553+113, PKS 2155-304, PKS 0426-380, PKS 0537-441 
and PKS 0301-243) have been reported to show year-like quasi-periodic variations (QPVs) with significance $>3\sigma$. As these findings are based on few cycles only, care needs to be taken to properly account 
for random variations which can produce intervals of seemingly periodic behaviour.
}
{We present results of an updated timing analysis for six blazars (adding PKS 0447-439 to the above), utilizing suitable methods to evaluate their longterm variability properties and to search for QPVs in
their light curves.
}
{We generate $\gamma$-ray light curves covering almost ten years, study their timing properties and 
search for QPVs using the Lomb-Scargle Periodogram and the Wavelet Z-transform. Extended Monte Carlo simulations are used to evaluate the statistical significance.
}
{ (1) Comparing their probability density functions (PDFs), all sources (except PG 1553+113) exhibit a 
clear deviation from a Gaussian distribution, but are consistent with being log-normal, suggesting that
the underlying variability is of a non-linear, multiplicative nature.
(2) Apart from PKS 0301-243 the power spectral density (PSD) for all investigated blazars is close to 
flicker noise (power-law slope $-1$).
(3) Possible QPVs with a local significance $\simgt 3 \sigma$ are found in all light curves (apart 
from PKS 0426-380 and PKS 0537-441), with observed periods between $(1.7-2.8)$ yr. The evidence is 
strongly reduced, however, if evaluated in terms of a global significance.
}
{ 
Our results advise caution as to the significance of reported year-like HE QPVs in blazars. 
Somewhat surprisingly, the putative, redhift-corrected periods are all clustering around $\sim 1.6$ yr. 
We speculate on possible implications for QPV generation.
}

   \maketitle
\section{Introduction}
Blazars belong to the most luminous and variable extragalactic sources in the Universe. They represent a special sub-class of AGN characterized by a relativistic plasma jet oriented very close to the line of sight at angles $\theta \leq 1/\Gamma$, where $\Gamma$ is the bulk Lorentz factor ( \citep[e.g.,][]{1995PASP..107..803U, 2015ApJ...810...14A}. Blazar sources are known to be variable across all wavelengths from radio frequencies to TeV $\gamma$-ray energies, and on a wide range of timescales from sub-minute to several years \citep[e.g.][]{2002ApJ...581..127B, 2007ApJ...664L..71A}.
Their multi-wavelength spectral energy distribution (SED) often exhibits two bumps. The first, low-energy bump (peaking at infrared to X-ray frequencies) is usually interpreted as synchrotron emission from highly relativistic electrons within in the jet. The second bump, on the other hand, peaking in the HE range has been frequently related to inverse Compton emission (IC) up-scattering of various soft photon fields (mostly synchrotron or external thermal radiation), although hadronic interactions may contribute as well.

Periodic variability in the light curves of blazars has been investigated extensively in the radio and optical band \citep[e.g.,][]{2002A&A...381....1F, 2006A&A...456L...1K, 
	2013MNRAS.436L.114K, 2016ApJ...832...47B}. In this context the two prominent sources OJ 287 and 3C 279 are worth mentioning, with longterm optical periods of $\sim 12$ and $29.6$ years being reported, respectively \citep{2006ApJ...646...36V,1538-3873-121-885-1172}. 
In the high-energy gamma-ray regime only five blazars (PG 1553+113, PKS 2155-304, PKS 0301-243, PKS 0426-380 and PKS 0537-441) have been reported so far exhibiting longterm quasi-periodic variability in their $\gamma$-ray fluxes with significance higher than $3\sigma$ \citep{Ackermann:2015wda, 2041-8205-793-1-L1, 2017ApJ...845...82Z, Zhang:2017ear, Sandrinelli:2015ijk, Prokhorov:2017amk}. 
Apart from PKS 0537-441, these QPVs have been seen over the whole length of the light curve.
In the majority of these cases, attempts have been made to relate the detected periodic behaviour at 
different wavelengths and on various timescales to the motion of two SMBHs in a binary black holes system, general helical jet structures, shocks or instabilities of the disk or jet-plasma flow \citep[e.g.,][]{2004ApJ...615L...5R, 2011JApA...32..147W, 2017MNRAS.465..161S}.

In the present paper we use Fermi-LAT $\gamma$-ray data between $08.2008$ and $12.2017$ to re-evaluate the five blazars mentioned above and to investigate an additional source, PKS 0449-439. Results of a periodicity search based on the generalized Lomb-Scargle periodogram (GLSP) and the Weighted Wavelet Z-transform (WWZ) are presented. The signifiance of 
the inferred periods is then estimated on the basis of Bootstrap resampling and light curves simulations using the Timmer$\&$ Koenig as well as the Emmanoulopoulos algorithm \citep{1995A&A...300..707T, 2013MNRAS.433..907E}

The paper is organized as follows. Section~2 provides an exemplary illustration of the performed Fermi-LAT data analysis procedures to generate $\gamma$-ray light curves based on a binned likelihood method and aperture photometry technique, respectively. In Sec.~3 we describe the GLSP and WWZ technique used here to search for periodicity.
Signifiance estimation and variability characterization is presented in Sec.~4.
Finally, Sec.~5 provides a summary and discussion of the results.

\section{Observations and Analysis}

\subsection{Fermi-LAT likelihood analysis}

Fermi-LAT on board the Fermi satellite is a electron-positron pair-conversion $\gamma$-ray detector sensitive to photon in energy range between $\sim 20$ MeV and $500$ GeV. The LAT has a wide field of view of $\sim2.4$ sr and observes the entire sky every 2 orbits. It has a point spread function (PSF) $<0.8^{\circ}$
and the largest effective area of approximately $8000$ cm$^{2}$ above 1 GeV. Fermi-LAT data were downloaded from 
the Fermi-LAT Data Server\footnote{https://fermi.gsfc.nasa.gov/ssc/data/access/} and events were selected between $2008$ August $4$ and $2018$ April $1$ (MET from $239557418.0$ to $541555205.0$), covering about $\sim 9.5$ years with energy range from $100$ MeV to $500$ GeV in a circular region of interest (ROI) of $15$ degree centered on the position of each source \citep{2013arXiv1303.3514A}.

A binned maximum likelihood analysis \citep{1996ApJ...461..396M} is performed using the Pass 8 (P8R2) algorithms and employing 
the standard Fermi Science Tools v10r0p5 software package. We follow the standard procedure provided by the Fermi Science 
Support Center (FSSC) to reduce the data. 
In addition, photons coming from zenith angles larger than $90$ degree were all rejected to reduce the background from gamma rays produced in the atmosphere of the Earth (albedo) with P8R2\_SOURCE\_V6 instrument response functions (IRF). We performed standard quality cuts in accordance with the Pass8 data analysis criteria. The background emission was modeled using the Galactic and isotropic diffuse emission gll\_iem\_v06\_iso and P8R2\_SOURCE\_V6\_v06 files two. All sources from the third LAT source catalog (3FGL; \citep{2015ApJS..218...23A}
within the ROI are included in the model to ensure a satisfactory background modeling.

A binned-likelihood analysis with a bin size of  $0.1^{\circ}$ was performed. The spectral indices were allowed to vary 
for sources located within a radius of $5^{\circ}$ around the position of the investigated source. The Test Statistic 
value (TS), defined as TS$= -2\ln(L_0/L_1)$ was used to determine the source detection significance, with threshold set to TS$=25$ ($\sim5\sigma$). The significance of a source detection is given by $\sim \sqrt[]{TS}\sigma$   \citep{2010ApJS..187..460A}. In the case of PKS 0447-439 for example, the likelihood analysis reveals a point source with a high statistical significance TS $\simeq 29305$ corresponding to $171\sigma$. Both energy and temporal bins with TS$< 9$ (or $\sim 3\sigma$) are set as upper limits throughout the paper. 

The appearance of new sources within the ROI could in principle strongly influence the $\gamma$-ray spectrum and light curve 
of the investigated source. In order to investigate this we also searched for possible new $\gamma$-ray sources within a FOV 
of $15^{\circ}$. Our analysis of $9.5$ year of Fermi-LAT data indicates additional transient sources beyond the 3FGL catalog;
these were appropriately taken into account in the full analysis, i.e. for each additional source we sequentially add a new point source with a standard spectral definition (PowerLaw) and maximize the likelihood as a function of its flux.

We produced the $\gamma$-ray SED and light curves of each sources through the binned maximum likelihood fitting technique, respectively with \textit{gtlike} to determine the flux and TS value for each time bin. The effect of energy dispersion below $300$ MeV is also accounted for in the analysis. For that we enabled the energy dispersion correction.
\begin{figure}[htb]
	\centering
    \includegraphics[width=0.49\textwidth]{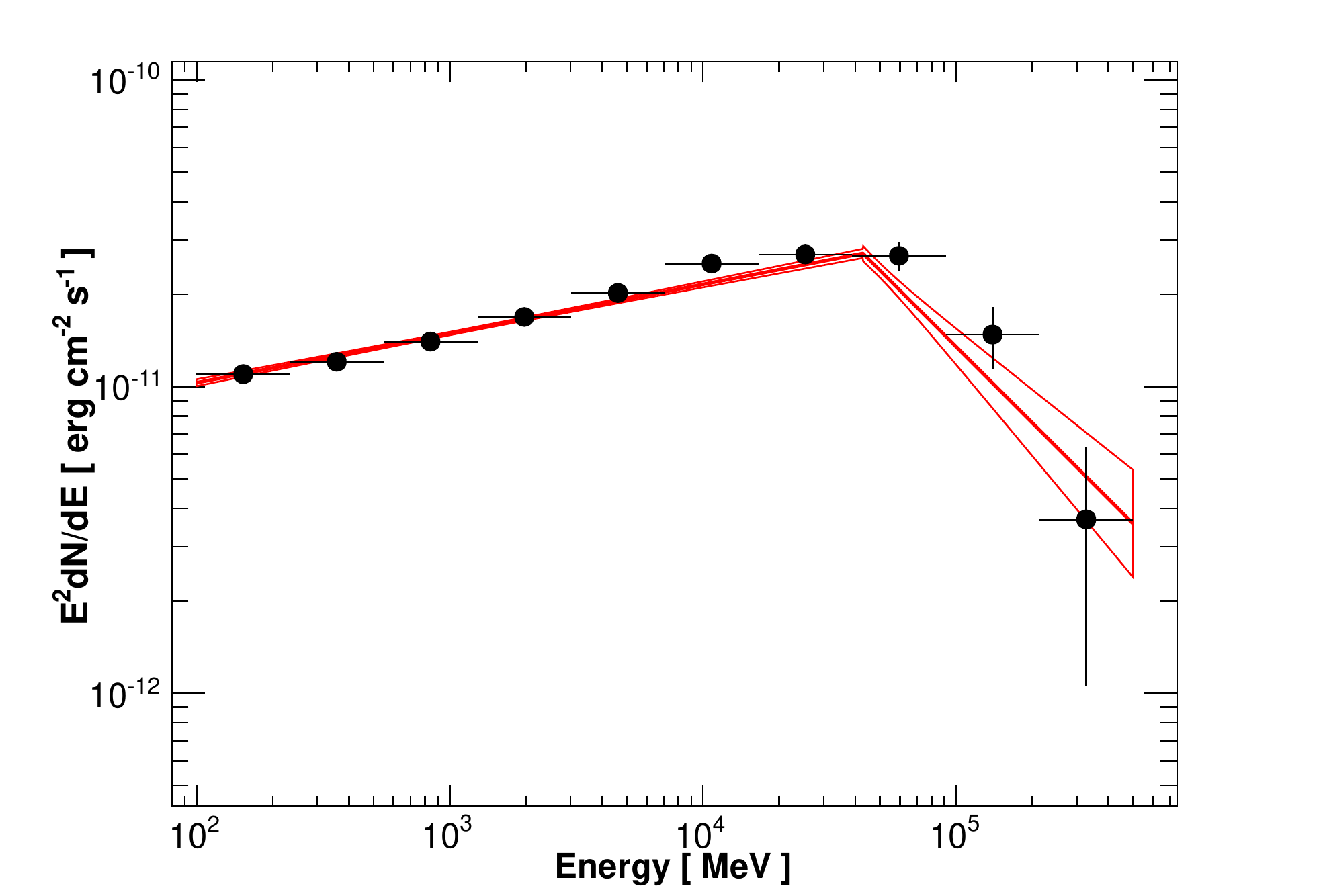} 
   \caption{The SED of PKS 0447-439 in the energy band of 100 MeV to 500 GeV as 
   			extracted from the complete data set (2008-2018) along with a Broken 
            Power Law fit (solid red line). The regions show the $1\sigma$ confidence 
            intervals resulting from the fit with BPL. The resulting break energy is 
            $E_{b} = 42.84 \pm 1.78$.}
   \label{fullSED}%
\end{figure}

As an example, the SED of PKS 0447-439 is shown in \textcolor{blue}{Fig.~1} for the full time and energy range using 10 energy bins. We performed maximum likelihood analyses, exploring different spectral forms for the whole energy range, namely Single Power-Law (PL), Broken Power-Law (BPL), Log-Parabola (LP) and Power-Law Super Exponential Cutoff (PLExpCutOff).
In the case of PKS 0447-439 a likelihood ratio test comparison yields TS$=-2 \log(L_{PL}/L_{BPL}) \simeq 53.8$, thus preferring BPL 
over PL at a significance level of $7.3\ \sigma$. The best fit indices are $\Gamma_1=1.84\pm 0.01$ and $\Gamma_2 = 2.82\pm0.17$ with a break at an energy $E_b = (42.8\pm1.8)$ GeV. The obtained results of the maximum likelihood analysis are summarized for the different models in the \textcolor{blue}{Table 1}. 
\begin{table*}
\centering
\begin{tabular*}{\textwidth}{@{\extracolsep{\stretch{1}}}*{8}{c}@{}}
  \toprule
  \toprule
  \centering{Model}  & Energy 	    	 & Spectral Parameters 	   	        &  N$_{pred.}$  &  Integral Flux 		   &    TS  	& Log-  	   & Log-Likelihood 	\\
  				     &  [GeV] 			 &				   	   		        & 			   & $10^{-8}[cm^{-2}s^{-1}]$  &  			&  Likelihood  & Ratio-Test	 	    \\ 
  \toprule 
  Power Law  (PL)        & $0.1$ .. $500$  	 & $\Gamma = 1.867 \pm 0.009 $      &  $16724$  & $(7.83 \pm 0.13)$  	   & $29293$  & -$402489.3$  & $-$						\\
  \midrule 
  Log Parabola 	     & $0.1$ .. $500$    & $\Gamma_{\alpha} = 1.742 \pm 0.025$       &  $15726$  & $(7.39 \pm 0.15)$ 	       & $29234$  & -$402479.8$  & $19.00$						\\
  	 (LP)            & 					 & $\Gamma_{\beta}  = 0.020 \pm 0.004$       &  	   &   					       &            &       	   & $(4.4 \mathlarger{\sigma})$	\\
  \midrule
  Broken PowerLaw    & $0.1$ .. $500$    & $\Gamma_{1} = 1.839 \pm 0.009$   &  $16234$  & $(7.62 \pm 0.13)$         & $29305$  & -$402462.4$  & $53.80$            \\
    (BPL)            & 				     & $\Gamma_{2} = 2.825 \pm 0.171$   &  			   & 					       &	        &         	   & $(7.3 \mathlarger{\sigma})$     \\
                     & 					 & $E_{b}      = 42.84 \pm 1.78$    &  			   &     				       & 		    &              &                    \\
  \midrule
  PLSuperExpCutOff 	 & $0.1$ .. $500$  	 & $\Gamma_{1}  = 1.818 \pm 0.012$  &  $16119$   & $(7.51 \pm 0.14)$        & $29282$  & -$402464.2$ & $ 50.20$             \\
                     & 					 & $\Gamma_{2}  = 1.031 \pm 0.195$  &      			   &      			       &  		    &			  & $(7.1 \mathlarger{\sigma})$  \\
                     & 					 & $E_{c}       = 137.9 \pm 24.3$   &      			   &     				   & 		    &             & \\
  \bottomrule                             
\end{tabular*}

\caption{Parameters obtained for the fit of the Fermi-LAT  energy spectrum of PKS 0447-439 using Power Law (PL), Log Parabola (LP), 
			Broken Power Law (BPL) and PLSuperExpCutOff (PLExp) spectral model. The last column gives the significance, obtained using Log-Likelihood 
            Test by comparing the log-Likelihood values for each model against those for the power-law model. Only statistical errors are shown. A BPL is preferred here.  N$_{pred.}$ represents the number of predicted photons.}
\end{table*}

\subsection{Fermi-LAT $\gamma$-ray light curves}
We generate $\gamma$-ray light curves for all our six sources using the individual best fit results (i.e., a BPL model in the case of 
PKS 0447-439), see \textcolor{blue}{Table~2} for details. For comparison two methods are employed: the maximum likelihood optimization 
and the aperture photometry.
In the analysis with the first method the $\gamma$-ray light curves are produced by using separate equal time-bins of one month (30 days). We apply a maximum likelihood fitting technique by running the ScienceTools \textit{}{gtlike} to extract the flux and TS value for each time bin. In this computationally more intensive procedure, the $\gamma$-events are selected in a circular ROI of $15^{\circ}$ radius centered on the position of the source. The resulting light curve for PKS 0447-439 is shown in \textcolor{blue}{Fig.~2} and has an average $\gamma$-ray flux of $\langle \mathlarger{\phi} \rangle = (7.62\pm 0.13)\times10^{-8}$cm$^{2}$s$^{-1}$. The source exhibits flux variability during the whole observational period, with flux levels of peak-to-peak oscillations changing by a factor four. 
The calculated fractional variability index is F$_{var}=(47.91 \pm 1.16)\%$ \citep{2003MNRAS.345.1271V, 2002ApJ...568..610E}. 

\begin{figure}[htb]
            \centering
            \includegraphics[width=0.49\textwidth]{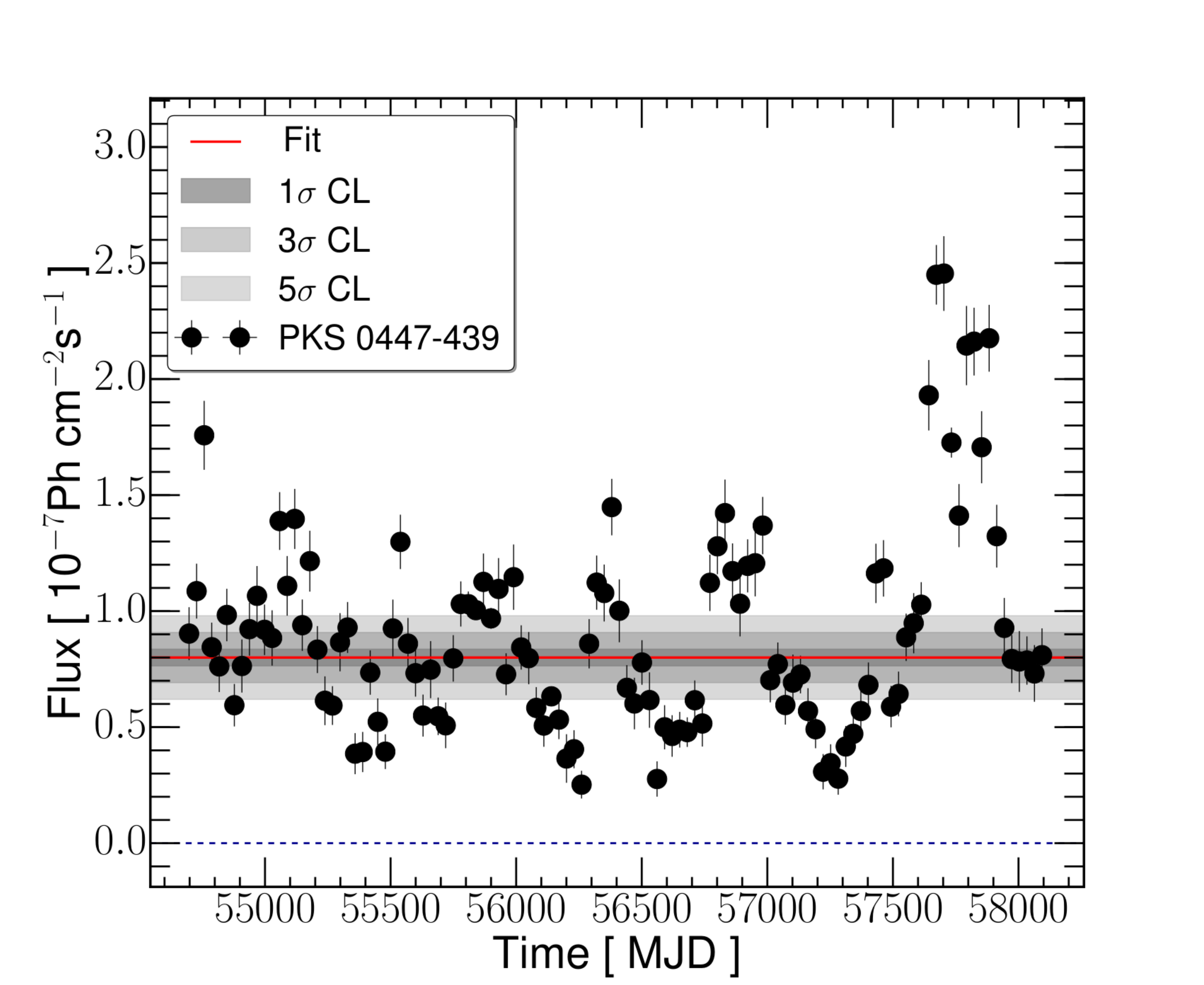}
			\caption{The $\gamma$-ray lightcurve of PKS 0447-439 from August 2008 to December 2017 in the energy range between 
                $100$ MeV and $500$ GeV in intervals of 30 days. The dotted line gives a fit 
                with a constant flux. There is clear evidence for variability. Vertical error bars indicate the one-sigma 
                error bars. 
                The gray shaded regions show the $1\sigma$,
                $3\sigma$ and $5\sigma$ confidence intervals resulting from the fit with a constant function.}
			\label{lc_Binned}
\end{figure}

\begin{figure}[htb]
        \centering
            \includegraphics[width=0.49\textwidth]{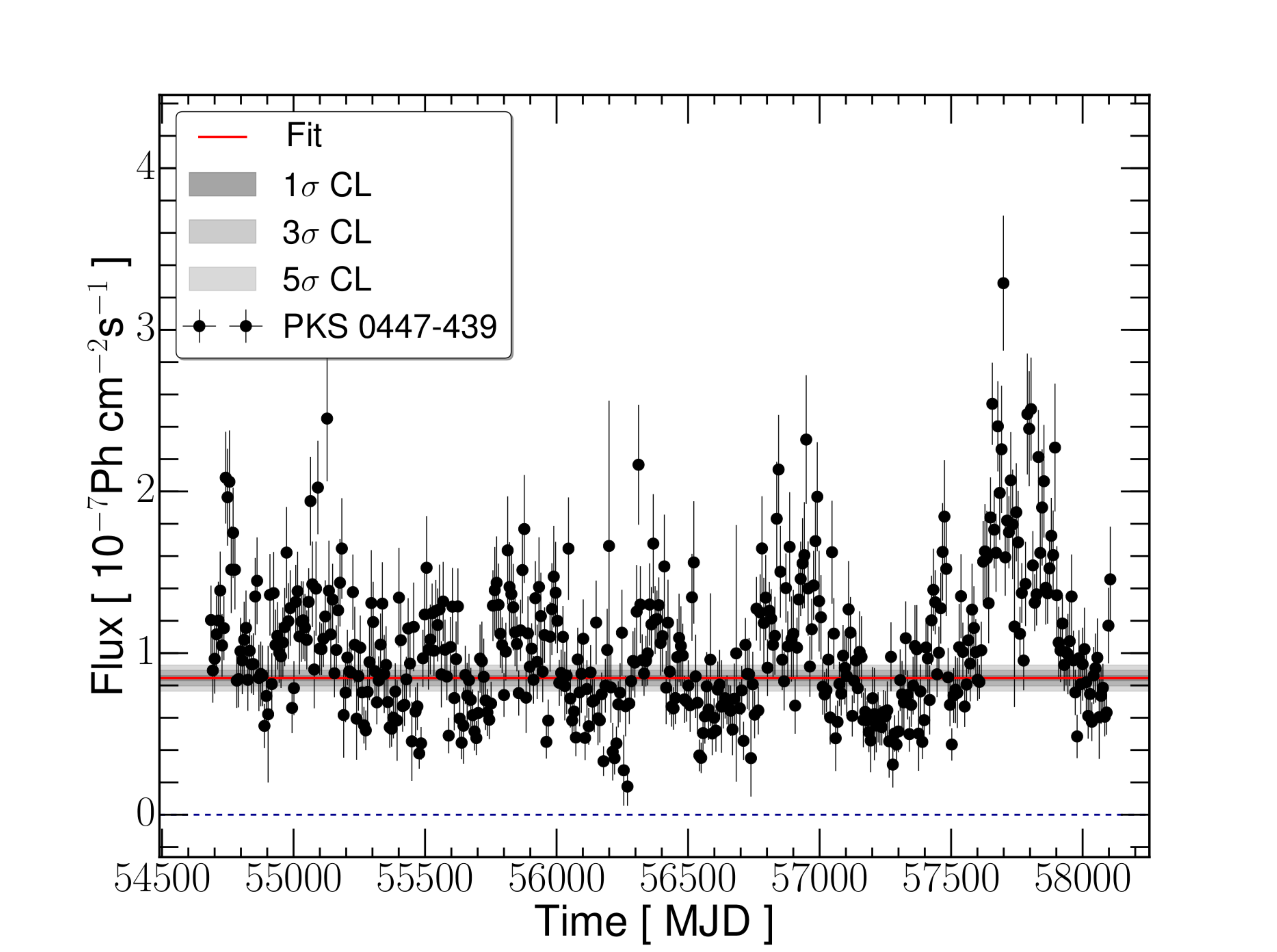}
			\caption{The $\gamma$-ray lightcurve of PKS 0447-439 from 2008 to 2018 in the energy range $(0.1 - 500)$ GeV 
             obtained by using the aperture photometry technique for a radius of $1^{\circ}$ around the source. A binning
             of one week is employed. Error bars correspond to $1\sigma$ error bars. The gray shaded regions show the $1\sigma$,
             $3\sigma$ and $5\sigma$ confidence intervals resulting from the fit with a constant function.}
			\label{fig:lc_Apperture}
\end{figure}

The aperture photometry method on the other hand, provides a model-independent measure of the $\gamma$-ray flux and is less 
computationally demanding. It also enables the use of short time bins whereas the maximum likelihood technique requires that 
time bins contain sufficient photons for the analysis. We select a very small aperture radius of $1^{\circ}$ to exclude most 
background $\gamma$-events and to focus on the events that are most likely associated with the source itself. We produce  
$\gamma$-ray light curves in the range between $100$ MeV and $500$ GeV using a weekly binning. The exposure of each time 
bin is determined with the ScienceTools \textit{gtexposure}. Figure ~\ref{fig:lc_Apperture} shows the resultant  
$\gamma$-ray light curve for PKS 0447-439 obtained with this method.

\begin{figure}[htb] 
			 \centering
			 \includegraphics[width=0.49\textwidth]{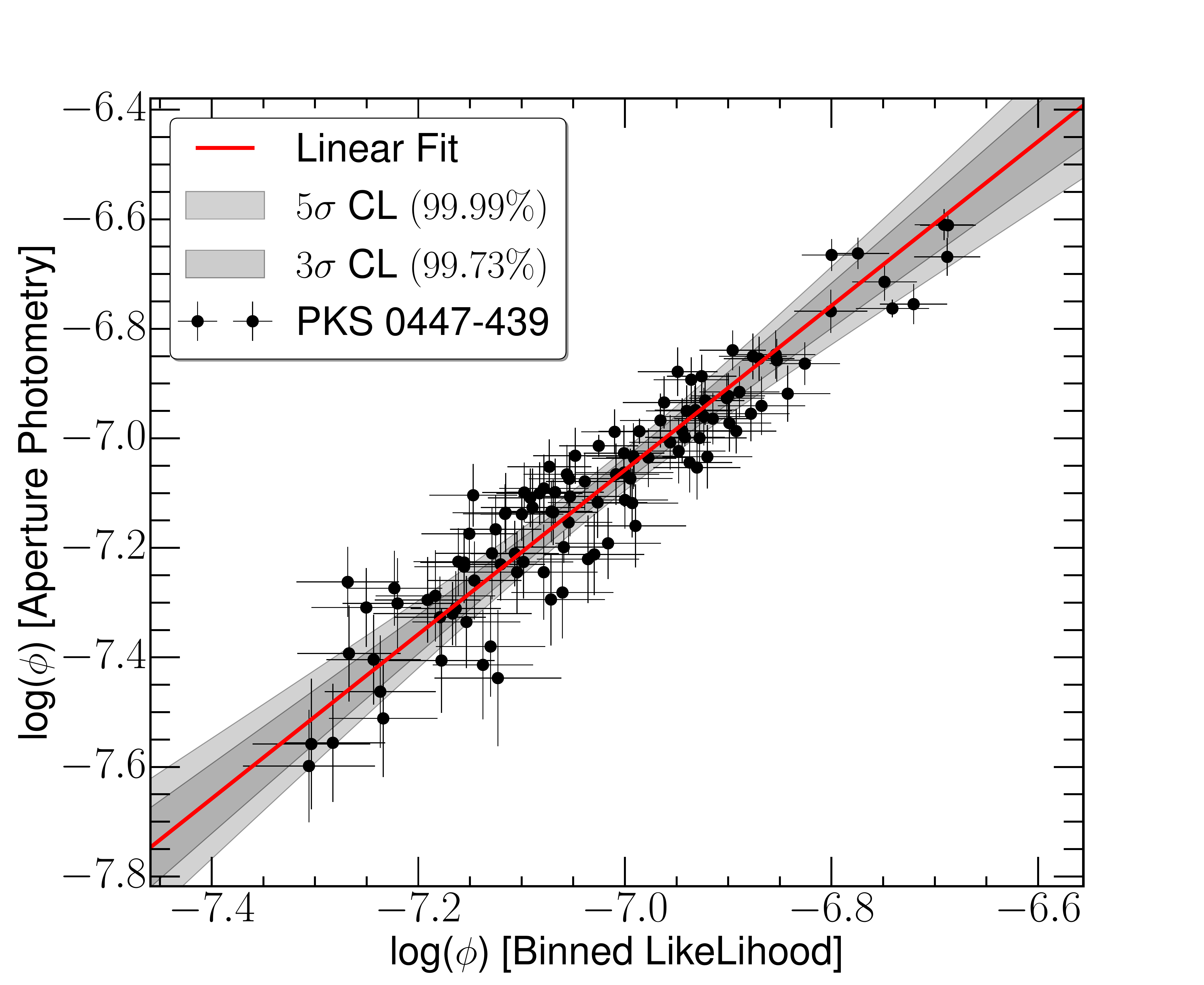}  
			 \caption{The Aperture Photometry flux in comparison with the binned likelihood flux calculated for PKS 0447-439 (logarithmic scale). 
             		   The red solid line represents a fit of the data with a linear function and the gray shaded regions show the resulting 
                       $3\sigma$ and $5\sigma$ confidence intervals. Error bars indicate the $1\sigma$ error interval.No background subtraction 
                       is used in the Aperture Photometry method.} 
			 \label{fig:AP vs BL}
\end{figure}

In order to verify that the aperture photometry (AP) approach gives results which are consistent with the results obtained using the binned likelihood (BL) analysis, we performed diagnostic tests on each of the investigated blazars. The BL method is the preferred procedure for most types of Fermi-LAT analyses. Taking for granted that BL flux values are indeed more accurately measured than the AP flux, we examine the correlation between the AP and BL calculated flux to assess the performance of the AP method. This correlation is plotted for PKS 0447-439 in Fig.~\ref{fig:AP vs BL} using monthly binned light curves and the calculated value of the Pearson correlation coefficient is $\rho \sim 0.936$. 

\begin{table*}
\centering
\begin{tabular*}{\textwidth}{@{\extracolsep{\stretch{1}}}*{8}{c}@{}}
  \toprule
  \toprule
 Source 	 & Model  			 & Energy 	   	    & Spectral Parameters 			 &  Integral Flux 		    &    TS     & Ratio-   		 &	F$_{var}$			\\
 			 & 				     &  [GeV] 			&				   	   		  	 & $10^{-8}[cm^{-2}s^{-1}]$ &   		& Test	    	 &	[\%] 				\\ 
  \toprule 
PKS 0447-439 & Broken PowerLaw  &  $0.1$ .. $500$  & $\Gamma_{1} = 1.839 \pm 0.009$  & $(7.62 \pm 0.13)$        & $29305.1$ & $7.3 \sigma$   & $(47.91 \pm 1.16)$ 	\\
  	          & 				&    			   & $\Gamma_{2} = 2.825 \pm 0.171$  & 			                &	        & 				 &  					\\
              & 				&	 			   & $E_{b}      = 42.84 \pm 1.78$   & 			                & 		    &                &  					\\  
  \midrule
PG 1553+113   & Log Parabola 	 & $0.1$ .. $500$  & $\Gamma_{\alpha} = 1.587 \pm 0.013$  & $(5.59 \pm 0.15)$ 	    & $34701.6$ & $11.01 \sigma$ &	$(22.87 \pm 1.14)$	\\
  	          &        			 & 				   & $\Gamma_{\beta}  = 0.045 \pm 0.005$  &   					    &           & 				 &						\\    
              &        			 & 				   & $E_{b}  = 1.49 \pm 0.95$			  &   					    &           & 				 &						\\      
   \midrule
PKS 2155-304  & PLSuperExpCutOff & $0.1$ .. $500$  & $\Gamma_{1}  = 1.749 \pm 0.154$ & $(12.82 \pm 0.22)$       & $59800.5$ & $9.63 \sigma$  & $(36.62 \pm 1.07)$   \\
              & 			     & 				   & $\Gamma_{2}  = 0.620 \pm 0.154$ &      					&      		&  				 &  					\\
              & 				 & 			       & $E_{c} =   110.57 \pm 37.9$   &      					&     		&             	 & 						\\
   \midrule
PKS 0426-380 & Log Parabola      & $0.1$ .. $500$  & $\Gamma_{\alpha} = 1.742 \pm 0.025$   & $(24.05 \pm 0.31)$ 	    & $92892.5$ & $21.19 \sigma$ &	$(62.39 \pm 0.01)$ \\
  	         &        			 & 				   & $\Gamma_{\beta}  = 0.020 \pm 0.004$   &   					    &           & 				 &						\\
  			 &        			 & 				   & $E_{b}  = 0.76 \pm 0.03$    	         &   					    &           & 				 &						\\      
             \midrule
PKS 0301-243 & Log Parabola 	 & $0.1$ .. $500$  & $\Gamma_{\alpha} = 1.622 \pm 0.043$   & $(3.23 \pm 0.11)$ 	    & $9639.4$  & $5.94 \sigma$  &	$(82.28 \pm 1.90)$ 	\\
  	         &        			 & 				   & $\Gamma_{\beta}  = 0.039 \pm 0.006$   &   					    &           & 				 &						\\
	         &        			 & 				   & $E_{b}  = 0.11 \pm 0.02$    	         	&   					    &           & 				 &						\\      
    \midrule
PKS 0537-441 & PLSuperExpCutOff  & $0.1$ .. $500$  & $\Gamma_{1}  = 1.662 \pm 0.026$ & $(19.38 \pm 0.20)$       & $63570.9$ & $16.78 \sigma$ & $(85.13 \pm 0.08)$   \\
             & 					 & 				   & $\Gamma_{2} = 0.267 \pm 0.009$  &      					&      		&  		 		 &  					\\
             & 					 & 				   & $E_{c} =   0.24 \pm 0.06 $ 	         &      					&     		&              	 & 						\\
  \bottomrule                             
\end{tabular*}
\label{table :Spectral Parameters}
\caption{The spectral parameters obtained for the best fit of Fermi-LAT data of each sources from the energy range between $100$ MeV and $500$ GeV using four different spectral 
models PowerLaw, Broken PowerLaw, LogParabola and PLSuperExpCutoff. The forelast column gives the significance, obtained by comparing the likelihood values for each models against 
those for the PowerLaw model using Log likelihood ratio test. The last column summarizes the fractional variability amplitude F$_{var}$ calculated form monthly light curves \citep{2003MNRAS.345.1271V, 2002ApJ...568..610E}. Error bars correspond to $1\sigma$ error; only statistical errors are shown.}
\end{table*}

\subsection{Gaussian vs Log-normal Flux Distributions}\label{Non-Gaussianity}
The almost continuous detection of the investigated sources over time bins of seven days provides a good opportunity to investigate whether the
probability density function (PDF) for the long-term $\gamma$-ray emission reveals any preference for a gaussian (normal) or a log-normal flux distribution. It is expected that such a feature offers an important clue for understanding the central engine of a blazer \citep[e.g.,][]{2005MNRAS.359..345U,Shah2018}. A log-normal distribution of $\gamma$-ray fluxes, for example, would suggest that the mechanism driving the variability is multiplicative in nature rather than additive. Such multiplicative behaviour might possibly be related to accretion disk fluctuations \citep{1997MNRAS.292..679L, 2006MNRAS.367..801A} or the particle acceleration process itself \citep{Shah2018}.
As a possible caveat we note that PDF of stochastic processes with power spectral density (PSD) steeper than index $-1$ can show deviations from normality owing to divergence of power at low frequencies \citep[e.g.,][]{2003MNRAS.345.1271V}.

\begin{figure}
	\centering	
			\subfigure[PKS 0447-439 ]{\includegraphics[width=0.49 \textwidth]{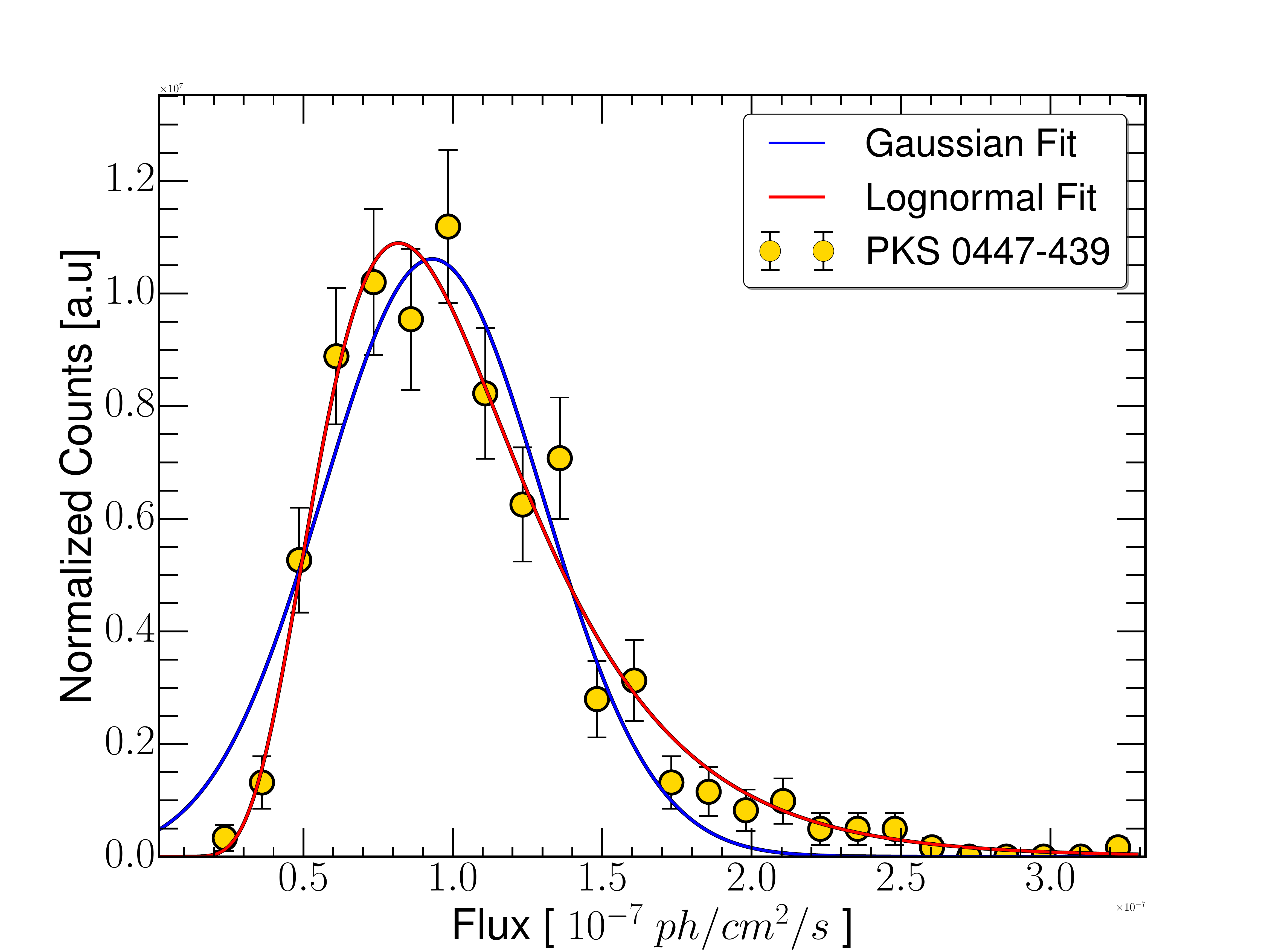}} 	
    		\subfigure[PG 1553+113  ]{\includegraphics[width=0.49\textwidth]{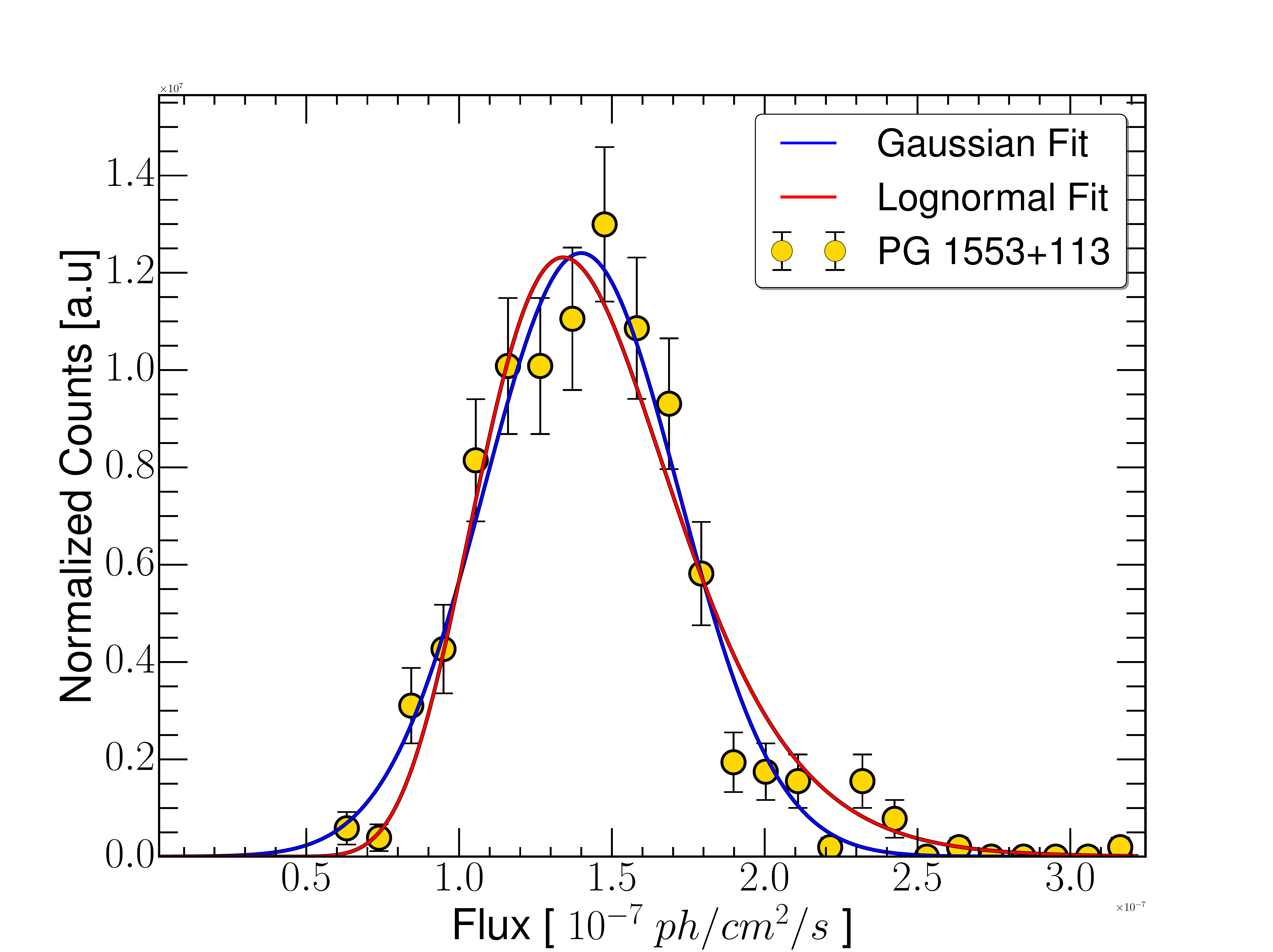}}  	 
            \subfigure[PKS 2155-304 ]{\includegraphics[width=0.49 \textwidth]{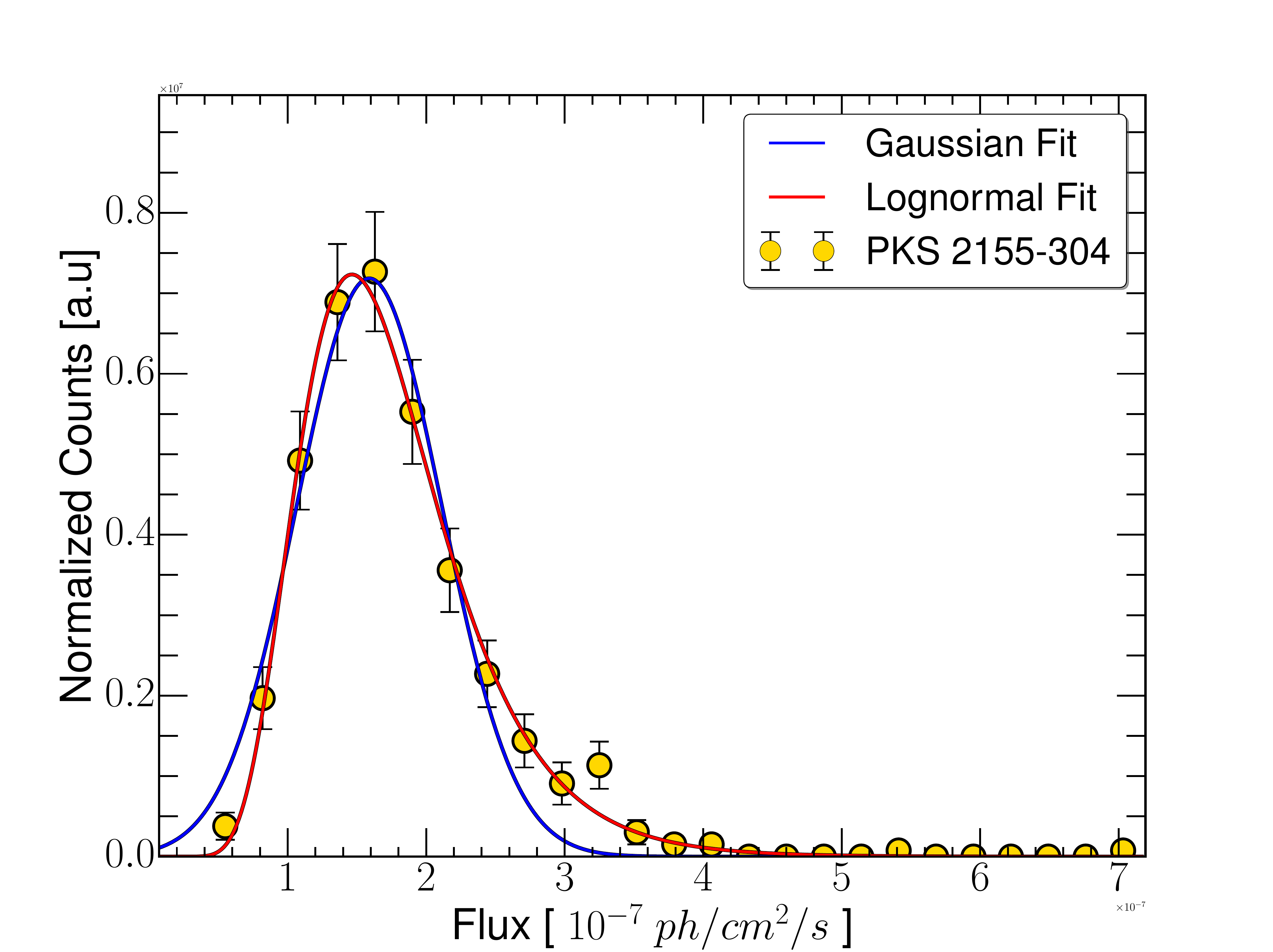}} 	
    		\caption{Normalized histograms of $\gamma$-ray photon fluxes fitted with a log-normal (solid red line) 
            and Gaussian (blue solid line), respectively. Apart from PG 1553+113 all sources show a clear preference
            for a log-normal distribution.}
			\label{fig:lognormality comparison}
\end{figure}

We explore the PDF and quantify this in terms of fluxes using flux-histograms. For each source the weekly $\gamma$-ray fluxes were distributed in a histogram of fluxes. We fit all histograms in log-scale, with Gaussian G($\phi$) and log-normal L($\phi$) distribution functions given by:
\begin{equation}
		\begin{aligned}
			L(\phi \textbf{ }| \textbf{ }\mu,\sigma)  = \frac{1}{\sqrt{2\pi}\sigma \phi} exp\textbf{ } \bigg(-\frac{\big(\log(\phi) -\mu \big)^2}{2\sigma^2}\bigg)  
		\end{aligned}
\end{equation}
and
\begin{equation}
		\begin{aligned}
    		G(\phi \textbf{ }| \textbf{ } \mu,\sigma)  = \frac{1}{\sqrt{2\pi}\sigma} exp\textbf{ } \bigg(-\frac{\big( \phi -\mu \big)^2}{2\sigma^2}\bigg)\,, 	
		\end{aligned}
\end{equation}
respectively, where $\sigma$ and $\mu$ are the standard deviation and the mean of the distribution, respectively. For illustration the obtained flux histograms for PKS 0447-439, PG 1553+113 and PKS 2155-304 are shown in Figure ~\ref{fig:lognormality comparison}. We compute the Anderson Darling test (AD) statistics \citep{ad1954} for each of the light curves as coming from a Gaussian or a log-normal distribution. The AD allows to test the null hypothesis that a sample is drawn from a population that follows a particular distribution, e.g. in our case a normal distribution. The results obtained are shown in Table~\ref{table:AD Test}. The obtained (AD Test) values provide strong evidence to reject the null hypothesis of an underlying normal distribution as the AD Test statistics is greater than the relevant critical value \citep[e.g.,][]{stephens1976}.

Except for PG 1553+113 all sources exhibit a clear deviation from a Gaussian distribution. The flux distributions are instead compatible with being log-normal, which suggests that the underlying variability is of a non-linear, multiplicative origin. Similar results are obtained from the D'Agostino's K$^2$ and the Shapiro-Wilk test, see Table~\ref{table:AD Test}.

\begin{table*}
  \centering
  \def\sym#1{\ifmmode^{#1}\else\(^{#1}\)\fi}
	\begin{tabular}{l*{9}{c}}
    \toprule
    \toprule
    \multicolumn{1}{c}{Source}       
	& \multicolumn{4}{c}{Log-normal }   & 
    \multicolumn{4}{c}{Gaussian } \\
    \cmidrule(lr){2-5}\cmidrule(lr){2-5}
    \cmidrule(lr){6-9}\cmidrule(lr){6-9}
 \multicolumn{1}{c}{Name} 	& \multicolumn{1}{c}{Center*}  & \multicolumn{1}{c}{Width} & \multicolumn{1}{c}{$\chi^2_{red.}$}  & \multicolumn{1}{c}{AD Test}  
							& \multicolumn{1}{c}{Center*}   & \multicolumn{1}{c}{Width}  & \multicolumn{1}{c}{$\chi^2_{red.}$}  & \multicolumn{1}{c}{AD Test} \\ 
 
							& \multicolumn{1}{c}{[ $\times10^{-7}$ ]}   & \multicolumn{1}{c}{[ $\times10^{-8}$ ]} &   & statistics 
							& \multicolumn{1}{c}{[ $\times10^{-8}$ ]}   & \multicolumn{1}{c}{[ $\times10^{-8}$ ]} &   & statistics \\ 
 \toprule 
 	PG  1553+113 & $(1.41 \pm 0.02)$ & $(2.13 \pm 0.13)$ & $14.59$ & $2.57$   & $(1.40 \pm 0.02)$ & $(3.21 \pm 0.15)$  & $12.29$ &  $1.83$  \\
    PKS 0447-439 & $(0.97 \pm 0.01)$ & $(2.48 \pm 0.15)$ & $5.70$  & $0.18$   & $(0.93 \pm 0.02)$ & $(3.66 \pm 0.16)$  & $9.94$  &  $6.61$  \\
    PKS 2155-304 & $(1.65 \pm 0.02)$ & $(3.65 \pm 0.23)$ & $8.00$  & $0.35$   & $(1.58 \pm 0.02)$ & $(5.14 \pm 0.22)$  & $14.38$ & $8.76$  \\
    PKS 0426-380 & $(2.19 \pm 0.09)$ & $(8.99 \pm 0.58)$ & $13.29$ & $0.79$   & $(1.91 \pm 0.08)$ & $(12.42 \pm 0.89)$ & $14.02$ & $7.84$  \\
    PKS 0301-243 & $(0.65 \pm 0.01)$ & $(1.57 \pm 0.09)$ & $6.56$  & $0.77$   & $(0.62 \pm 0.01)$ & $(2.37 \pm 0.08)$  & $17.52$ &  $22.52$  \\
    PKS 0537-441 & $(1.28 \pm 0.04)$ & $(4.62 \pm 0.29)$ & $16.30$ & $11.13$  & $(1.02 \pm 0.09)$ & $(7.82 \pm 0.94)$  & $35.99$ & $26.41$  \\
    \bottomrule
  \end{tabular}
  \caption{The table shows the best Fit parameters for the log-normal and normal (Gaussian) distribution, respectively; the AD Test statistics and $\chi^2_{red.}$ are the Anderson Darling test and the reduced $\chi^2$, respectively. [*] in units of photon/cm$^{2}s^{-1}$.} 
  \label{table:AD Test}
\end{table*}

\begin{table}
\centering
\begin{tabular}{|l|ll|ll|}
	\hline
    \hline
	Source 			& K$^{2}$    & P$_{value}$ 			 & Shapi.-  & P$_{value}$		\\
    	   			& Stati.	& ($\chi^{2}$ Prob.)	 & 	Wilk	&	(Prob.)			\\
    \hline
		PG 1553+113  & $56.8$   &  $4.6\times10^{-13}$  & $0.972$  & $4.7\times10^{-8}$  \\
		PKS 0447-439 & $101.3$  &  $1.0\times10^{-22}$  & $0.935$  & $9.9\times10^{-14}$ \\
        PKS 2155-304 & $222.1$  &  $5.8\times10^{-49}$  & $0.893$  & $6.4\times10^{-18}$ \\
        PKS 0426-380 & $90.7$   &  $2.0\times10^{-20}$  & $0.927$  & $1.2\times10^{-14}$ \\
        PKS 0301-243 & $721.4$  &  $2.2\times10^{-157}$ & $0.607$  & $8.1\times10^{-32}$ \\
     	PKS 0537-441 & $185.4$  &  $5.6\times10^{-41}$  & $0.815$  & $2.9\times10^{-23}$ \\
	\hline
\end{tabular}
		\caption{D'Agostino's K$^2$ Test tests the null hypothesis that a sample comes from a normal distribution. 
        It is based on D’Agostino and Pearson’s test \citep{doi:10.1093/biomet/57.3.679, doi:10.1093/biomet/60.3.613} 
        and combines skew and kurtosis to produce an omnibus test of normality. Similarly, the Shapiro-Wilk test quantifies how likely it is that the data was drawn from a Gaussian distribution 
        \citep{10.2307/2333709}}
		\label{table : normality test}
\end{table}

\section{Searching for periodicity}

In order to search for periodicity in blazars we apply two widely used methods, the Lomb-Scargle Periodogram and the Weighted Wavelet Z-transform to light curves.
\subsection{Lomb-Scargle Periodogram}

The Lomb-Scargle periodogram (LSP) is a commonly used algorithm for detecting and characterizing periodicity in unevenly sampled light curves \citep{1976Ap&SS..39..447L,1982ApJ...263..835S}. The standard normalized Lomb-Scargle Periodogram (LSP) is equivalent to fitting sine waves of the form $y(t)= A\textbf{ }\cos(\omega t) + B\textbf{ }\sin(\omega t)$, and is defined for a uneven simpled time series ($t_{i},y_{i}$ ) as
\begin{eqnarray}
		\mathlarger{P(\omega)} &=&  \dfrac{ \Bigg [ \mathlarger{\sum}_{i} y_{i} \textbf{ } \cos\textbf{ }\big (\omega (t_{i}-\tau)\big ) 
                            	        \Bigg]^{2} }{ 2 \mathlarger{\sum_{i}} \cos^{2}\big 	(\omega(t_{i}-\tau)\big )} +    
                                     	\frac{ \Bigg [  \mathlarger{\sum_{i}} y_{i} \textbf{ } \sin\textbf{ }\big (\omega (t_{i}-\tau)\big ) \Bigg]^{2} }{ 			  										  								\mathlarger{2 \sum_{i}} \sin^{2}\big (\omega(t_{i}-\tau)\big )}   	\nonumber \\                                        
\mathlarger{\tan(\tau)} &=&  \frac{1}{2\omega} \dfrac{\mathlarger{\sum} \sin(2\omega t_{i})}{\mathlarger{\sum_{i}} \cos(2\omega t_{i})}\,. 		
\end{eqnarray}
The standard LSP, however, has some limitations. On the one hand, it does not take measurement errors into account; on the other hand, in the calculation the mean of the flux is subtracted, which assumes that the mean of the data and the mean of the fitted sine function are the same. To overcome those limitations we instead use the generalized LSP 
(GLSP) \citep{2009A&A...496..577Z}. As an example the calculated GLSP for PKS 0447-439 using monthly $\gamma$-ray light curves (generated based on the binned likelihood) is shown in Fig.~\ref{fig: PKS 0447-439 GLSP 50K Simulation}.
A strong peak at around a period of $945 \pm 40 $ days is apparent in this figure. We have estimated the uncertainty of the calculated period based on the half-widths at half-maxima (HWHM) of Gaussian fits to the profile at the position of the highest peak. An evaluation of confidence levels determined from Monte Carlo simulations of colored noise is explained in detail in Sec. 4.

\begin{figure}[H]
				\centering
                \includegraphics[width=0.49\textwidth]{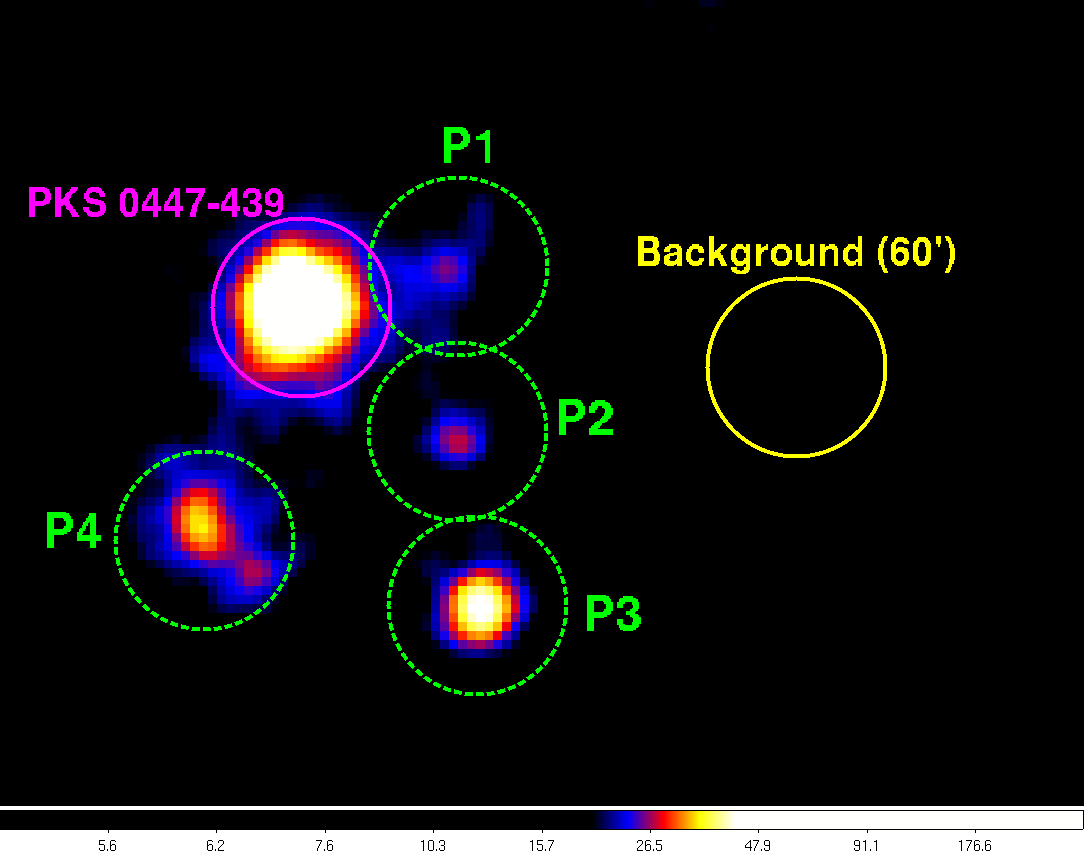}
				\caption{Smoothed count map (logarithmic scale) of the PKS 0447-439 region in energy range from  
                         $100$ MeV up to $500$ GeV as seen by Fermi-LAT based on data between 
                         August 2008 and December 2017. The color bar has units of counts per 
                         pixel and the pixel dimensions are $0.1\times0.1$ degrees. The locations 
                         of the background region and surrounding sources (labeled P1 = new source beyond the 3FGL catalog, 
                         P2 = 3FGL J0438.8-4519, P3 = 3FGL J0437.2-4713 and 
                         P4 = 3FGL J0455.7-4617.) are shown as circles. The magenta circle represents the location of 
                         PKS 0447-439 and the yellow circles the background region.}
				\label{fig:ROI CountMap}
\end{figure}
\begin{figure}[H]
				\centering
				\includegraphics[width=0.5\textwidth]{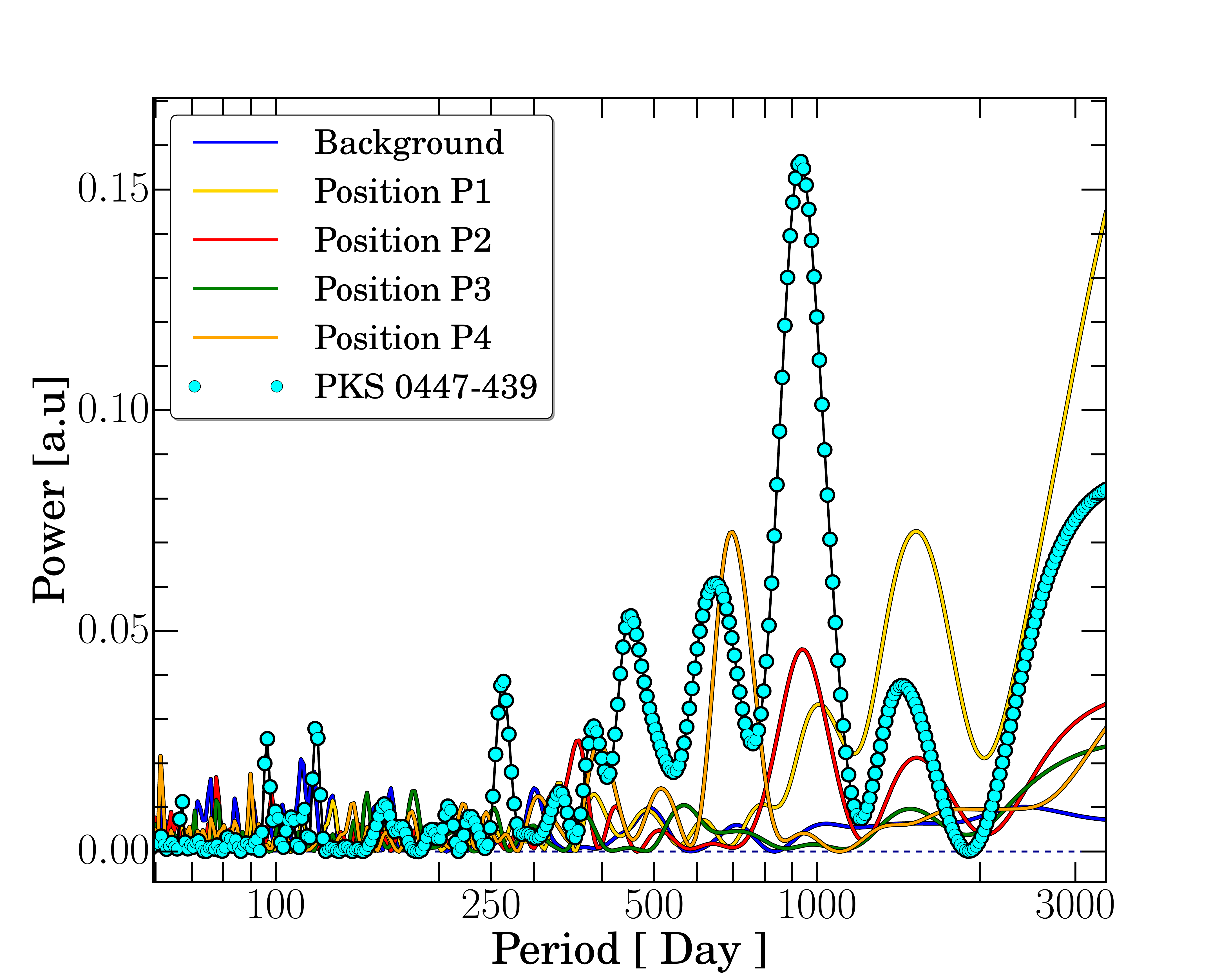}
				\caption{Generalized Lomb-Scargle periodograms 
                obtained based on light curves generated using aperture photometry for 
                PKS 0447-439 and surrounding sources (P1,P2,P3,P4) shown as circles in 
                figure~\ref{fig:ROI CountMap}.}
				\label{BG_LombScargle}
\end{figure}
\begin{figure*} 
				\centering
                 \includegraphics[width=0.80\textwidth]{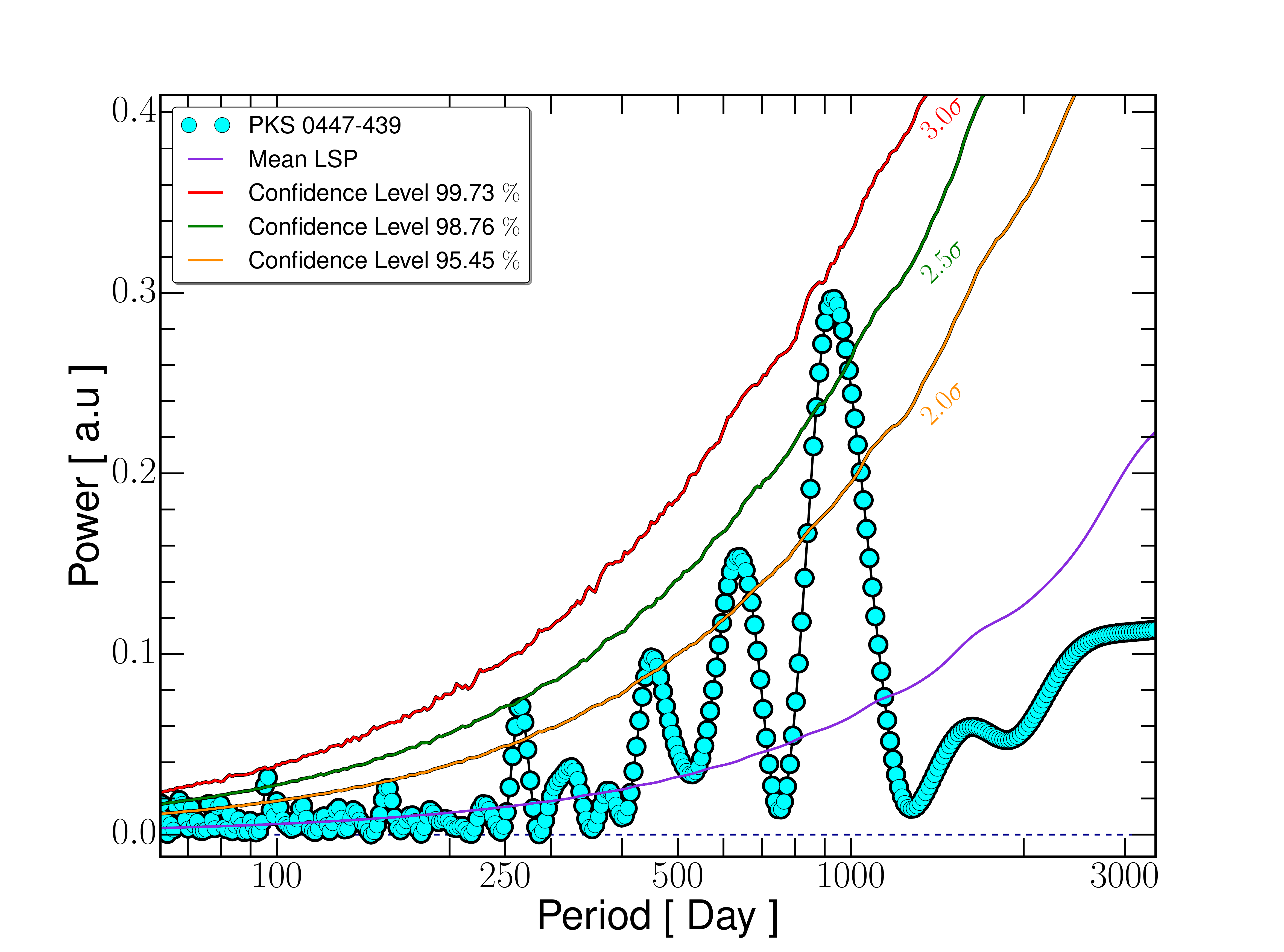}
    			\caption{The Generalized Lomb-Scargle periodogram (GLPS) obtained using the monthly light curves of PKS 0447-439 is shown in cyan (circles); the red solid line, green line and orange line represent the $3\sigma$, $2.5\sigma$, $2\sigma$ 
                         confidence level, respectively calculated based on simulations of $50000$ light curves using 				
                		 \citet{1995A&A...300..707T} method.}
				\label{fig: PKS 0447-439 GLSP 50K Simulation}
\end{figure*}

For comparison the GLSP method has also been applied to the light curves based on aperture photometry.
The possible interference of nearby sources (including background contribution) is evaluated for each source 
as exemplary shown in Fig.~\ref{fig:ROI CountMap} and Fig.~\ref{BG_LombScargle} for the case of PKS 0447-439. 
The results agree well with those obtained based on the binned likelihood method, as evident from Table~\ref{table:intrisic period}. 

The observed period P$_{\rm obs.}$ for a blazar at redshift $z$ is related to its intrinsic period 
by $P_{\rm int.} = P_{\rm obs.}/(1 + z)$. Table ~\ref{table:intrisic period} shows the GLSP results for 
all sources, except PKS 0537-441, based on both, monthly (generated based on the binned likelihood) 
and weekly (aperture photometry) binned light curves. Year-type HE periodicity is
found for all five sources, with the results based on aperture photometry 
and binned likelihood method being in good agreement. For the blazar
PKS 0537-441 no significant period has been found when searched for the whole light curve. Furthermore the source is estimated to have a red-noise power spectrum \citep{2019MNRAS.482.1270C}, which suggests a greater possibility of detecting a fake peak. This source has 
thus not been included in the following. We confirm, however, that during the initial $\sim 3$-year high 
state a peak can be seen at periodicity timescale of $\sim 280$ d  \citep[see e.g.,][]{Sandrinelli:2015ijk}. However, for this to be a robust result, a longer observation window showing a larger number of cycles is needed given the timescale of periodicity \citep{2016MNRAS.461.3145V}. 

\begin{table*}
  \centering
  \def\sym#1{\ifmmode^{#1}\else\(^{#1}\)\fi}%
	\begin{tabular}{l*{7}{c}}
    \toprule
    \toprule  
    \multicolumn{1}{c}{Source} & \multicolumn{1}{c}{Energy}	&  \multicolumn{1}{c}{Redshift}	 & \multicolumn{2}{c}{Observed Period }   &   \multicolumn{2}{c}{Intrinsic Period } \\
    \cmidrule(lr){4-5}\cmidrule(lr){4-5}
    \cmidrule(lr){6-7}\cmidrule(lr){6-7}
  & \multicolumn{1}{c}{[ GeV ] }& \multicolumn{1}{c}{$z$} &\multicolumn{1}{c}{Monthly bins*} & \multicolumn{1}{c}{Weekly bins*}  & \multicolumn{1}{c}{Monthly bins*} & \multicolumn{1}{c}{Weekly bins*} \\ 
 \toprule
 	PG  1553+113 &  $0.1$ .. $500$ &  $0.360$  & ($821.9 \pm 74.5$)    &   ($811.9 \pm 58.4$)   & ($604.3 \pm 54.8$) &   ($596.9 \pm 42.9$) \\
    PKS 0447-439 &  $0.1$ .. $500$ &  $0.343$  & ($929.0 \pm 88.4$)    &   ($935.4 \pm 81.0$)   & ($691.7 \pm 65.8$) &   ($696.5 \pm 60.3$) \\
    PKS 2155-304 &  $0.1$ .. $500$ &  $0.116$  &  ($619.7 \pm 40.8$)   &   ($626.7 \pm 35.6$)   & ($555.3 \pm 35.8$) &   ($561.5 \pm 31.9$) \\
    PKS 0426-380 &  $0.1$ .. $500$ &  $1.111$  & ($1218.0  \pm 158.7$) &   ($1225.8 \pm 146.9$) & ($576.9 \pm 75.2$) &   ($580.7 \pm 69.6$) \\
    PKS 0301-243 &  $0.1$ .. $500$ &  $0.260$  & ($755.1 \pm 58.4$)    &   ($760.5 \pm 90.5$)   & ($599.3 \pm 46.3$) &   ($603.6 \pm 71.8$) \\
    \bottomrule
  \end{tabular}
  	\caption{Periodicity results derived from the GLSP method for the monthly and
    weekly $\gamma$-ray light curves, respectively, and including the intrinsic 
    (redshift-corrected) period. [*] in units of days.} 
	\label{table:intrisic period}
    \end{table*}
    
\subsection{Weighted Wavelet Z-transform}
The GLSP provides an excellent tool for the periodicity analysis of light curves with unevenly-spaced sampling.
Nevertheless it does not account for the possibility that in some astrophysical systems quasi-periodic oscillations 
may develop that vary significantly in frequency and amplitude over a specified period of time. In such cases, 
the Weighted Wavelet Z-transform (WWZ) method turns out to be a more convenient technique for detecting and 
quantifying such variations \citep{1996AJ....112.1709F, doi:10.1190/geo2012-0199.1}, and has been applied for 
the timing analysis of AGN light curves at different wavelengths \citep[e.g.][]{2016ApJ...832...47B, 2015ApJ...805...91M}. 
The method is based on a similar concept like LSP, where sinusoidal functions are used to fit the data; however, 
the waves can now be localized in both time and frequency domain to account for the possible transient nature of 
the QPOs \citep{1998BAMS...79...61T, 2014A&A...568A..34B}.

The WWZ is based on a weighed projection onto three trial functions, $y(t) = \sum_{i} y_{i}\phi_{i}(t)$
\begin{equation}
 \text{ $y(t) = \sum_{i} y_{i}\phi_{i}(t)$ :}\qquad	
 \left\{
 \begin{aligned}
 		\phi_{1}(t) = 1 (t) = 1 \\
 		 \phi_{2}(t) =  cos\textbf{ }\big(\omega(t-\tau)\big) \\
        \phi_{3}(t) =  sin\textbf{ }\big(\omega(t-\tau)\big) 
	\end{aligned}
	\right.
\end{equation}
where the 'best-fit' coefficients $y_i$ are the coefficients for which the model function $y(t)$ best fits the 
data. For each projection the statistic weight is given by 
\begin{equation}
\begin{aligned}
	\mathlarger{\omega_{i} = e^{-c\omega(t_{i}-\tau)^2}}\,,
\end{aligned}
\end{equation}
where $c$ is a constant that determines how rapidly the Morlet wavelet decays, and is usually chosen to be close to 
$0.0125$. The WWZ power can then be finally defined as
\begin{equation}
WWZ = \frac{(N_{eff}-3)V_{y}}{2(V_{x} -V_{y})}\,,
\end{equation} with V${_x}$ and V${_y}$ as weighted variations of the data and the model function, respectively 
and N$_{eff}$ representing the effective number of the data points (for more details, see \citep{1996AJ....112.1709F}. 

\begin{figure}[H]
		\centering
		\includegraphics[width=0.48\textwidth]{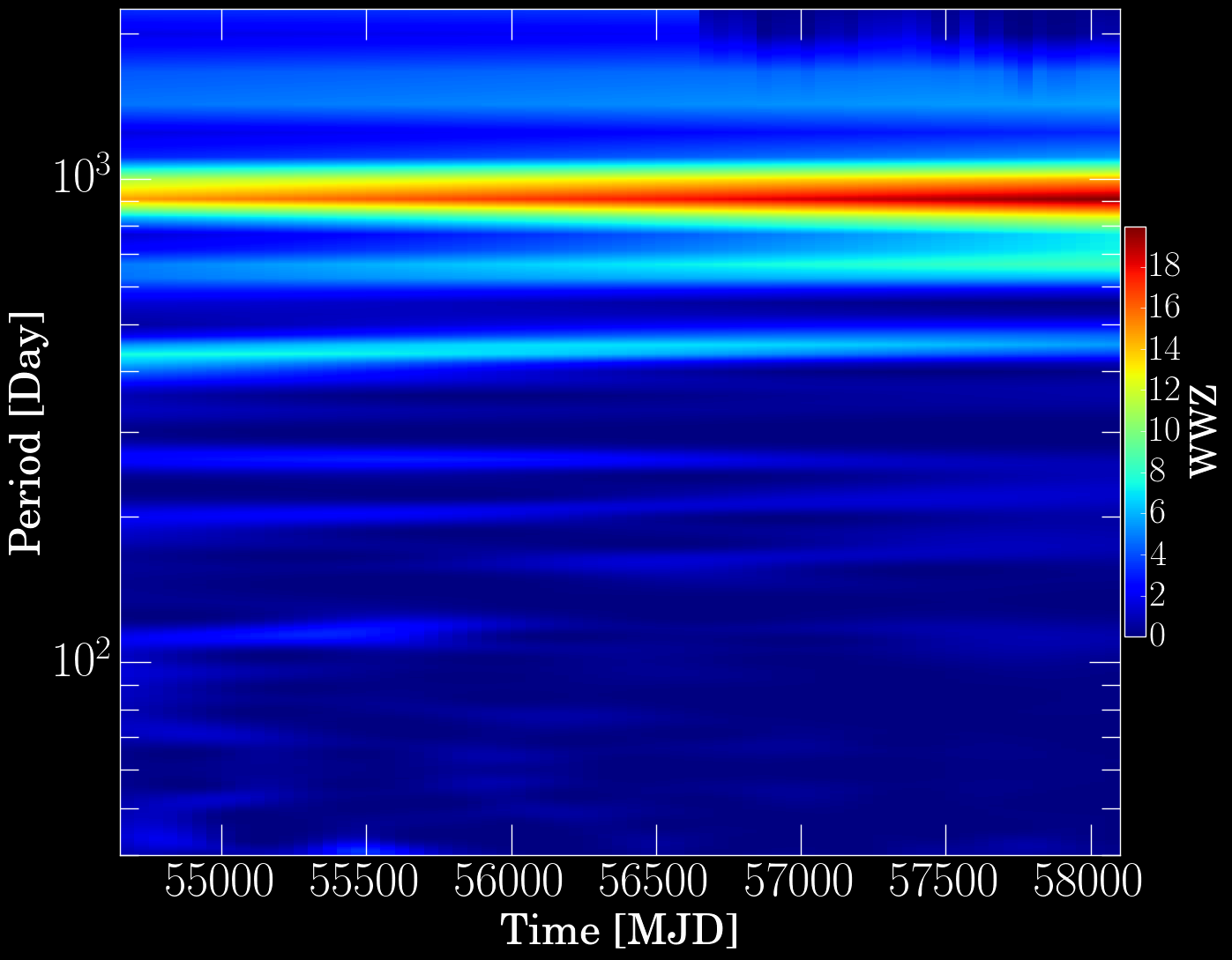}
		\caption{Two-dimensional Weighted Wavelet Z-transform of the $\gamma$-ray 
                light curve of PKS 0447-439 based on Fermi-LAT data from August 2008 
                - December 2017. The color scale represents the Z-statistics of 
                the WWZ power of a certain period at a given time. 
                The panel shows the signature of a quasi-periodic oscillation at a 
                (observed) period of $\sim2.6$ years without any significant changes 
                over time.}
		\label{fig:wwz}
\end{figure}
The WWZ powers for PKS 0447-439 are showed in Fig.~\ref{fig:wwz} as a function of both observing time (horizontal axis) 
and period (vertical axis). The peaks in the power characterize the strength and duration of a possible quasi-periodic modulation 
in the data. The WWZ indicates a characteristic period at $(937.3 \pm 152.7)$ days (around $2.6$ years) with 
peak power of about $18.5$ and no significant changes over time. The constancy of the period indicates that the 
quasi-periodicity is probably driven by a physical process that is stable over duration of observation. The results of 
the WWZ method are shown in Table~\ref{table : wwz period}. The results agree well with the GLSP results.
\begin{table}
\centering
\begin{tabular}{lccc}
	\hline
    \hline
	Source 	Name	  		& $z$		& Observed Period   	& Intrinsic  Period	   \\
    	   				&			&   [ days ]			& [ days ]	 		   \\
    \hline
		PG 1553+113  	& $0.360$  	& $807.8  \pm 102.4$  &  $593.9 \pm 75.3$   \\
		PKS 0447-439 	& $0.343$ 	& $937.3  \pm 152.7$ 	&  $697.9 \pm 113.7$  \\
        PKS 2155-304 	& $0.116$ 	& $624.7  \pm 69.5$ 	&  $559.8 \pm 62.3$   \\
        PKS 0426-380 	& $1.111$ 	& $1257.9 \pm 280.4$  &  $595.9 \pm 132.8$  \\
        PKS 0301-243 	& $0.260$ 	& $752.4  \pm 89.2$ 	&  $597.1 \pm 70.8$   \\
	\hline
\end{tabular}
		\caption{Summary results of the strongest periods for the Weighted Wavelet Z-transform  
        		method WWZ based on the monthly $\gamma$-ray light curves, including  
             	observed and intrinsic period. 
        		} 
		\label{table : wwz period}
\end{table}

\section{Significance and Uncertainty Estimation } 
The effects related to irregular sampling of a light curve and the noisy nature of the periodogram can in some 
situation lead to the generation of false (artificial) periods in the GLSP that could be mistaken as real periodic 
signal of the source. For this reason, it is important to take such effects into accounts when searching for 
periodicities, and to analyze their impact on GLSP carefully.\\ 
The variability of AGN light curves often exhibits
a colored-noise-like behavior with power spectral density (PSD) characterized by a simple power-law of the form 
PSD($\nu$) $\sim \nu^{-\beta}$ where $\nu$ is the (temporal) frequency and $\beta$ the power-law index. 
 
 \subsection{Power Spectrum Response Method (PSRESP)}
\begin{figure}[H]
\centering	
			\subfigure[Aperture Photometry]{\includegraphics[width=0.5\textwidth]{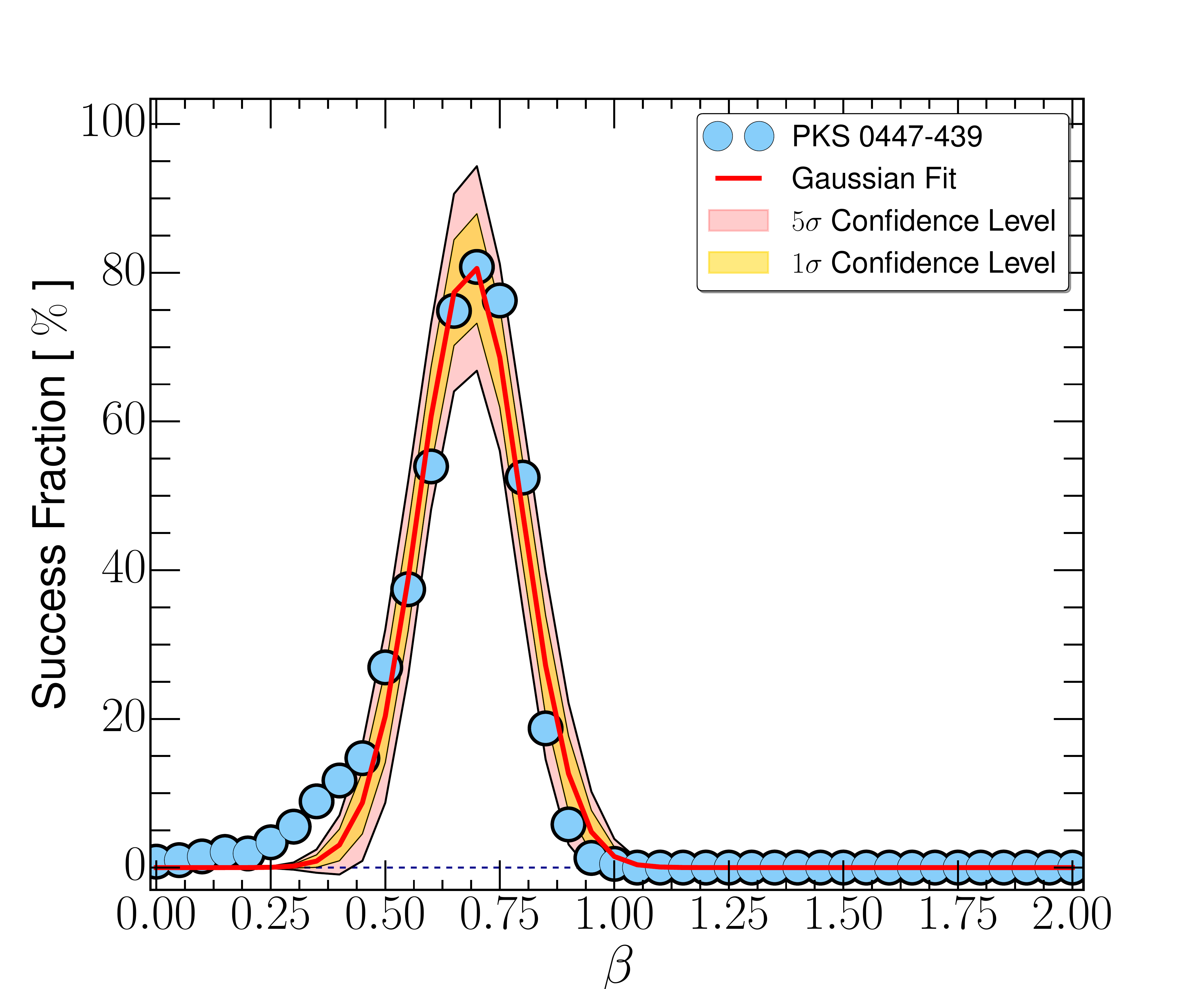}}
			\subfigure[Binned Likelihood  ]{\includegraphics[width=0.5\textwidth]{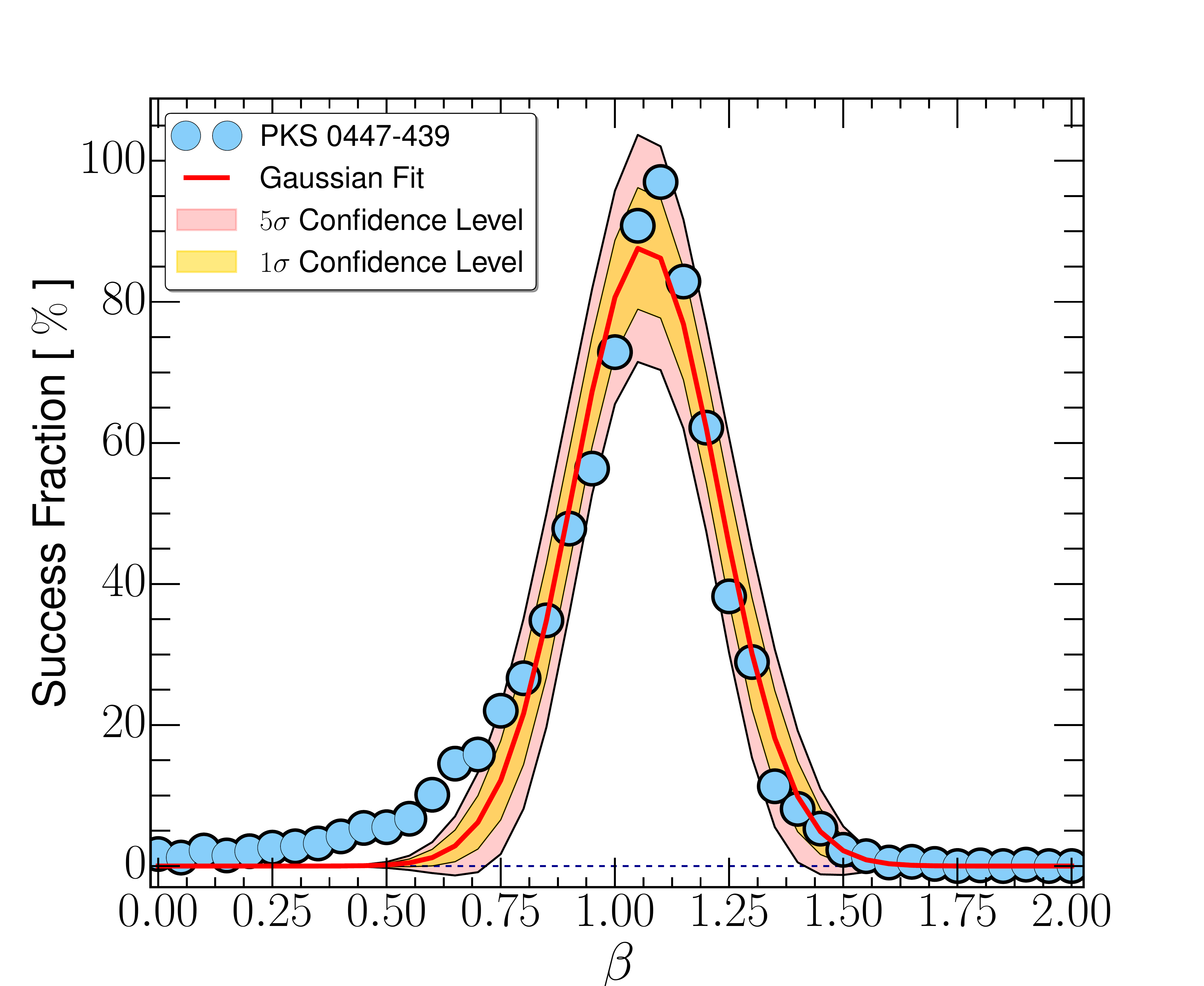}}
    		\caption{Success fraction  vs. slope ($\beta$) for all the three PSDs (Fermi-LAT). The success fractions 
            		 indicate the goodness of fit obtained from the PSRESP method (see text) .
                     (a) Monthly (b) Weekly light curves}
			\label{fig:success fraction}
\end{figure}

In order to determine the appropriate form of the underlying colored noise needed as an input for simulating the 
light curves, we first applied the Power Spectrum Response method (PSRESP; \citet{2002MNRAS.332..231U}), which is 
a widely used technique for the characterization of AGN power spectra \citep[e.g.,][]{2008ApJ...689...79C, 2016ApJ...832...47B}. 
The method attempts to fit the binned periodogram with different realisations of a given 
PSD model in order to estimate the model which maximizes the probability that observed PSD can be reproduced.
Our implementation of the PSRESP method is described in full detail in \citet{2008ApJ...689...79C}. Selected
details are as follow: 

We consider a simple power-law model for the underlying power spectrum of the form
\begin{equation}
 	P(\nu) = A\bigg(\frac{\nu}{\nu_{0}} \bigg)^{-\beta} + C_{noise}\,, 
\end{equation}
where $A$ is the amplitude of the model at the reference frequency $\nu_{0}$, $\beta$ correspond 
to the power-spectral slope and $C_{noise}$ is a constant which is fixed at the Poisson noise level for the light 
curve. We simulate $N=1000$ light curves starting from the underlying model and using the Timmer \& Koenig algorithm 
(see below), and re-sample these with the same sampling interval of the observed light curves. 
We then calculate the periodogram of the observed light curve (P($\nu$)$_{obs.}$) and that of each of the 
simulated light curves (P($\nu$)$_{sim,i}$ where $i=1,N$). The $\chi^{2}_{dist,i}$ statistic is calculated 
from the underlying model average and observed PSD of each light curve, with
\begin{eqnarray}
		\mathlarger{\chi^{2}_{dist,i}} = \mathlarger{\sum\limits_{\nu=\nu_{min}}^{\nu_{max}}} \frac{\big [ P(\nu)_{sim,i} - \langle P(\nu)_{sim} \rangle \big]^2 }{ \langle \Delta P(\nu)_{sim} \rangle^2}\,,  \\ 
     \   
        \mathlarger{\chi^{2}_{obs}} = \mathlarger{\sum\limits_{\nu=\nu_{min}}^{\nu_{max}}} \frac{\big [ P(\nu)_{obs} -\langle P(\nu)_{sim} \rangle \big]^2 }{ \langle \Delta P(\nu)_{sim} \rangle^2}\,, 
\end{eqnarray}
where $\nu_{min}$ and $\nu_{max}$ are the minimum and maximum frequencies measured by P($\nu$)$_{obs.}$, respectively. 
We calculate the $\chi^{2}_{dist}$ for each simulated PSD P($\nu$)$_{sim,i}$ over all frequencies between $\nu_{min}$ 
and $\nu_{max}$ with respect to the data. We then compare $\chi^{2}_{dist}$ with the $\chi^{2}_{obs}$ distribution and 
count the number $m$ of $\chi^{2}_{dist,i}$ for which  $\chi^{2}_{obs}$ is smaller than $\chi^{2}_{dist,i}$.
Finally, we calculate the success fraction m/M (goodness of fit), which gives the probabilities of a model being accepted, 
for a range of $\beta$ from $0.0$ to $2.0$ with step size of $0.05$. 
The obtained results according to the PSRESP method are summarized in Table ~\ref{table:psd slope} and shown in 
Fig. ~\ref{fig:success fraction}
\begin{table*}
  \centering
 \def\sym#1{\ifmmode^{#1}\else\(^{#1}\)\fi}%
	\begin{tabular}{l*{6}{c}}
    \toprule
    \toprule
    \multicolumn{1}{c}{Source}	&  \multicolumn{2}{c}{Simulation Method} & \multicolumn{2}{c}{Monthly}  & \multicolumn{2}{c}{Weekly} \\
       \cmidrule(lr){2-3}\cmidrule(lr){2-3}
       	\cmidrule(lr){4-5}\cmidrule(lr){6-7}
       	\cmidrule(lr){4-5}\cmidrule(lr){6-7}
  &	\multicolumn{1}{c}{PSRESP}& \multicolumn{1}{c}{Simulation} & \multicolumn{1}{c}{PSD-Slope} & \multicolumn{1}{c}{Success Fraction} & 
      \multicolumn{1}{c}{PSD-Slope} & \multicolumn{1}{c}{Success Fraction} \\
  & \multicolumn{1}{c}{Model} & \multicolumn{1}{c}{Algorithm} & \multicolumn{1}{c}{$\beta_{m}$} & \multicolumn{1}{c}{[ \% ]} & 
      \multicolumn{1}{c}{$\beta_{w}$} & \multicolumn{1}{c}{[ \% ]} \\   
    \midrule
 	PG 1553+113  & Power Law  & Timmer \& Koenig    & $0.88 \pm 0.13$    &	$96.0 \pm 3.2$	 & $ 0.66 \pm 0.11$ 	&	$82.4 \pm 3.5$	  \\
     			 & Power Law  & Emmanoulopoulos     & $0.90 \pm 0.14$    &	$93.0 \pm 2.9$	 & $ 0.67 \pm 0.11$ 	&	$84.8 \pm 3.6$	  \\	
   \addlinespace
   PKS 0447-439  & Power Law    & Timmer \& Koenig 	& $1.06 \pm 0.15$    &	$98.4 \pm 3.0$	 & $ 0.68 \pm 0.10$ 	&	$88.3 \pm 3.8$	  \\       
   	              & Power Law	& Emmanoulopoulos   & $1.07 \pm 0.13$    &	$83.5 \pm 5.2$	 & $ 0.69 \pm 0.10$ 	&	$86.6 \pm 4.1$	  \\	   
   \addlinespace
   PKS 2155-304   & Power Law   &  Timmer \& Koenig & $0.95 \pm 0.13$     &	$97.4 \pm 3.1$	 & $ 0.88 \pm 0.09$ 	&	$86.9 \pm 3.3$	  \\    
    			  & Power Law	& Emmanoulopoulos   & $0.99 \pm 0.14$     &	$94.4 \pm 2.7$	 & $ 0.89 \pm 0.09$ 	&	$90.5 \pm 3.6$	  \\	
   \addlinespace
   PKS 0426-380   & Power Law   & Timmer \& Koenig   & $0.98 \pm 0.22$    &	$95.1 \pm 5.0$	 & $ 1.08 \pm 0.08$ 	&	$96.5 \pm 3.6$	  \\     
    			  & Power Law   & Emmanoulopoulos    & $1.02 \pm 0.23$    &	$88.6 \pm 4.1$	 & $ 1.09 \pm 0.09$ 	&	$94.2 \pm 3.7$	  \\	
   \addlinespace
   PKS 0301-243   & Power Law   &  Timmer \&  Koenig & $ 0.32 \pm 0.20$   & $96.9 \pm 3.6$	 & $ 0.53 \pm 0.14$ 	&	$94.8 \pm 5.3$	  \\  
   				  & Power Law   & Emmanoulopoulos    & $ 0.36 \pm 0.22$   &	$95.1 \pm 2.9$	 & $ 0.58 \pm 0.15$ 	&	$93.9 \pm 3.2$	  \\	
	\bottomrule
  \end{tabular}
  	\caption{The table shows the slopes $\beta$ obtained from fitting the $\gamma$-ray power spectra with simple power-law model, and the 
    correspondents success fraction calculated based on the PSRESP method \citep{2002MNRAS.332..231U} (see subsection \textcolor{blue}{$4.1$}) 
    (the success fractions indicate the goodness of fit obtained from the PSRESP method). The $\beta$ values have been determined using both 
    monthly and weekly light curves, as well as the Timmer \& Koenig (1995) and Emmanoulopoulos et al. (2013) simulation methods. The errors 
    are results from fitting Gaussians to each of the slope profile.}\label{table:psd slope}
\end{table*}
\subsection{MC-simulations of colored-noise light curves}
The Timmer and Koenig (TK) algorithm \citep{1995A&A...300..707T} is a commonly used method to produce artificial light curves. This technique allows to generate 
non-deterministic (stochastic Gaussian-distributed) time series from a given underlying PSD model by randomizing both the phase and the 
amplitude of the Fourier components. Limitations could arise, however, for light curves that exhibit strong deviations from Gaussian 
distributions (e.g., a burst-like behaviour). 

Given the preference for log-normality, we  also simulate $5\times10^{4}$
light curves using the method proposed by \citep[][E13]{2013MNRAS.433..907E} to obtain best-fitting results from the PSRESP method.
The latter E13 method is able to account for a general PDF, i.e. to match both the PSD and the probability density function (PDF) of an observed 
light curve, thus relaxing restrictions of the TK method. This does in 
fact better comply with our previous findings of non-Gaussianity in Sec.~\ref{Non-Gaussianity}.

The results are shown in Fig.~\ref{table: significance} using the (main value of the) best fit slope $\beta$. In many cases the detected 
periods are close to or above $3\sigma$. Given available data there is rather little difference between the results based on the 
Timmer \& Koenig (1995) and the Emmanoulopoulos et al. (2013) methods. The outcome is, however, obviously dependent on the slope $\beta$. 
Within the PSRESP inferred range quite different results can be obtained as shown in Table~\ref{table: significance dependence}, suggesting 
that a narrowing-down of the PSD slope will be most relevant for assessing the real significance of the detected periods. 
\begin{table*}[!h]
  \centering
  \def\sym#1{\ifmmode^{#1}\else\(^{#1}\)\fi}%
	\begin{tabular}{l*{8}{c}}
    \toprule
    \toprule
       \multicolumn{1}{c}{Source}	&  \multicolumn{1}{c}{Simulation} & \multicolumn{3}{c}{Monthly}  & \multicolumn{3}{c}{Weekly} \\
       	\cmidrule(lr){3-5}\cmidrule(lr){3-5}
       	\cmidrule(lr){6-8}\cmidrule(lr){6-8} 
  	&	\multicolumn{1}{c}{Algorithm}& 
    \multicolumn{1}{c}{PSD-Slope} & \multicolumn{1}{c}{local} & \multicolumn{1}{c}{global } & 
    \multicolumn{1}{c}{PSD-Slope} &	\multicolumn{1}{c}{local} & \multicolumn{1}{c}{global } \\
 			&					& 
    \multicolumn{1}{c}{$\langle \beta_{m} \rangle$}	& \multicolumn{1}{c}{ Significance } & \multicolumn{1}{c}{Significance} & 
    \multicolumn{1}{c}{$\langle \beta_{w} \rangle$} & \multicolumn{1}{c}{ Significance } & \multicolumn{1}{c}{Significance} \\
    \midrule
	 PG 1553+113  & Timmer \& Koenig  & $ 0.88 $ & $3.89$ $\sigma$ & $2.14$ $\sigma$  & $ 0.66 $ & $3.70$ $\sigma$  & $2.07$ $\sigma$  \\
 			 	  & Emmanoulopoulos   & $ 0.90 $ & $3.82$ $\sigma$ & $1.73$ $\sigma$  & $ 0.67 $ & $3.83$ $\sigma$  & $2.16$ $\sigma$  \\
    \addlinespace
     PKS 0447-439  & Timmer \& Koenig & $ 1.06 $ & $2.87$ $\sigma$ & $1.00$ $\sigma$  & $ 0.68 $ & $3.26$ $\sigma$ & $1.49$ $\sigma$  \\
 			 	   & Emmanoulopoulos  & $ 1.07 $ & $2.94$ $\sigma$ & $1.16$ $\sigma$  & $ 0.69 $ & $3.16$ $\sigma$ & $1.46$ $\sigma$  \\
    \addlinespace
     PKS 2155-304  & Timmer \& Koenig & $ 0.95 $ & $3.62$ $\sigma$ & $1.61$ $\sigma$  & $ 0.89 $ & $3.32$ $\sigma$ & $0.84$ $\sigma$  \\  
 			 	   & Emmanoulopoulos  & $ 0.99 $ & $3.72$ $\sigma$ & $1.64$ $\sigma$  & $ 0.88 $ & $3.16$ $\sigma$ & $0.77$ $\sigma$  \\
    \addlinespace
     PKS 0426-380  & Timmer \& Koenig & $ 0.98 $ & $2.73$ $\sigma$ & $1.04$ $\sigma$  & $ 1.08 $ & $2.17$ $\sigma$ & $0.52$ $\sigma$  \\
 			 	   & Emmanoulopoulos  & $ 1.02 $ & $2.72$ $\sigma$ & $0.96$ $\sigma$  & $ 1.09 $ & $2.35$ $\sigma$ & $0.18$ $\sigma$  \\
    \addlinespace
     PKS 0301-243  & Timmer \& Koenig & $ 0.32 $ & $3.54$ $\sigma$ & $1.77$ $\sigma$  & $ 0.53 $ & $3.39$ $\sigma$ & $1.02$ $\sigma$  \\
 			 	   & Emmanoulopoulos  & $ 0.36 $ & $3.72$ $\sigma$ & $2.17$ $\sigma$  & $ 0.58 $ & $3.43$ $\sigma$ & $1.43$ $\sigma$  \\
   \bottomrule
  \end{tabular}
		\caption{The table shows the significance values of the observed periods obtained for the two method local and global significance evaluated using 
        		 the Timmer \& Koenig (1995) and the Emmanoulopoulos et al. (2013) and based on on the monthly and weekly $\gamma$-ray light curves. 
                 the slopes $\beta$ are obtained from fitting the $\gamma$-ray power spectra with simple power-law model, and the corresponding success 
                 fraction are calculated based on the PSRESP method \citep{2002MNRAS.332..231U}. 
         }
		\label{table: significance}
\end{table*}

\subsection{Statistical Confidence of GLSP-detected Period}
\subsubsection{Local significance}
In presence of an priori expectation of periodicity either from theory or from entirely independent observations, one can compute 
the statistical significance of the detected period at that position. This is the so-called "local significance" as referenced in 
several papers \citep[e.g.,][]{2011MNRAS.411..402B}. Thus, in the local method, the frequency channels corresponding to the 
GLSP-detected period were searched and the power spectra were recorded in which the peak power at this period exceeds the observed 
value. Finally, the fraction of occurrences with greater power than detected period is the probability of false positive, resulting 
from random red noise in the observed light curve.

\subsubsection{Global significance }  
In the majority of the cases of reported periodicities, particularly in gamma-rays, we do not have strong apriori indications of 
the expected periodicity. In such a circumstance, it is statistically more rigorous to evaluate the "global" rather than the local significance \citep[e.g.][]{2011MNRAS.411..402B}.
This constitutes computing the fraction of occurrences of larger peaks at any period within a reasonable range of timescales 
(dependent on the cadence properties of the light curve), relative to the detected one. This ensures that we factor in false-positives from this larger range of timescales rather than a specific period. 
This, so called "look elsewhere effect" (or also 
Multiple Testing Problem in Statistics), can be quantified in terms of a trial factor \citep[cf.][]{Lyons:1900zz, 2010EPJC...70..525G}.
This is defined as the ratio between the probability of detecting a peak (or period/excess) at some fixed frequency, to the 
probability of detecting it anywhere in the (tested) range. The "look elsewhere effect" has been explored and factored in 
for detection of resonant peaks in particle physics and indeed other fields including astronomy. We therefore re-evaluated 
the probability that the power of any observed peak is equal to or greater than a selected value somewhere in the periodogram 
and calculated in this wise the global significance \citep[cf.][]{2010MNRAS.402..307V}. The results are shown in 
Table.~\ref{table: significance} and reveal that in the absence of other physical reasons for restricting the period range the QPV evidence 
is strongly reduced, with none of the sources reaching $3\sigma$ significance. 
These findings are in line with similar indications in \citep{2019MNRAS.482.1270C}.
\begin{table*} 
  \centering
   \def\sym#1{\ifmmode^{#1}\else\(^{#1}\)\fi}%
	\begin{tabular}{l*{8}{c}}
    \toprule
    \toprule
     \multicolumn{1}{c}{Source}	&  \multicolumn{1}{c}{Simulation} & \multicolumn{3}{c}{Monthly}  & \multicolumn{3}{c}{Weekly} \\
       	\cmidrule(lr){3-5}\cmidrule(lr){3-5}
       	\cmidrule(lr){6-8}\cmidrule(lr){6-8}
  	&	\multicolumn{1}{c}{Algorithm}& 
    \multicolumn{1}{c}{PSD-Slope} & \multicolumn{1}{c}{local} & \multicolumn{1}{c}{global} & 
    \multicolumn{1}{c}{PSD-Slope} &	\multicolumn{1}{c}{local} & \multicolumn{1}{c}{global} \\
 	&	& 
    \multicolumn{1}{c}{$\beta_{m}$}	& \multicolumn{1}{c}{Significance} & \multicolumn{1}{c}{Significance} & 
    \multicolumn{1}{c}{$\beta_{w}$} & \multicolumn{1}{c}{Significance} & \multicolumn{1}{c}{Significance} \\
    \midrule
	 PG 1553+113  & Timmer \& Koenig & $ \beta_{min}               = 0.75 $ & $5.33$ $\sigma$ & $2.46$ $\sigma$  & $\beta_{min}               = 0.55 $ & $5.32$ $\sigma$ & $3.06$ $\sigma$  \\  %
 			 	  &  				 & $ \langle \beta_{m} \rangle = 0.88 $ & $3.89$ $\sigma$ & $2.14$ $\sigma$  & $\langle \beta_{w} \rangle = 0.66 $ & $3.70$ $\sigma$ & $2.07$ $\sigma$  \\
                  &  				 & $ \beta_{max}               = 1.01 $ & $3.72$ $\sigma$ & $1.86$ $\sigma$  & $ \beta_{max}              = 0.77 $ & $2.90$ $\sigma$ & $0.93$ $\sigma$  \\
     \addlinespace
     			  & Emmanoulopoulos  & $ \beta_{min}               = 0.76 $ & $5.33$ $\sigma$ & $2.40$ $\sigma$  & $ \beta_{min}              = 0.56 $ & $5.32$ $\sigma$ & $3.10$ $\sigma$  \\
                  &  				 & $ \langle \beta_{m} \rangle = 0.90 $ & $3.89$ $\sigma$ & $2.06$ $\sigma$  & $\langle \beta_{w} \rangle = 0.67 $ & $3.83$ $\sigma$ & $2.16$ $\sigma$  \\
                  &  				 & $ \beta_{max}               = 1.04 $ & $3.48$ $\sigma$ & $1.75$ $\sigma$  & $ \beta_{max}              = 0.78 $ & $2.79$ $\sigma$ & $0.79$ $\sigma$  \\ 
   \midrule
    PKS 0447-439  & Timmer \& Koenig & $ \beta_{min}               = 0.91 $ & $3.12$ $\sigma$ & $1.06$ $\sigma$  & $\beta_{min}               = 0.58 $ & $3.48$ $\sigma$ & $2.01$ $\sigma$  \\  %
 			 	  &  				 & $ \langle \beta_{m} \rangle = 1.06 $ & $2.87$ $\sigma$ & $1.00$ $\sigma$  & $\langle \beta_{w} \rangle = 0.68 $ & $3.26$ $\sigma$ & $1.49$ $\sigma$  \\
                  &  				 & $ \beta_{max}               = 1.21 $ & $2.67$ $\sigma$ & $0.46$ $\sigma$  & $ \beta_{max}              = 0.78 $ & $2.78$ $\sigma$ & $1.02$ $\sigma$  \\
     \addlinespace
     			  & Emmanoulopoulos  & $ \beta_{min}               = 0.94 $ & $3.11$ $\sigma$ & $1.04$ $\sigma$ & $ \beta_{min}              = 0.59 $ & $3.54$ $\sigma$ & $1.99$ $\sigma$  \\
                  &  				 & $ \langle \beta_{m} \rangle = 1.07 $ & $2.94$ $\sigma$ & $1.16$ $\sigma$  & $\langle \beta_{w} \rangle = 0.69 $ & $3.16$ $\sigma$ & $1.46$ $\sigma$  \\
                  &  				 & $ \beta_{max}               = 1.20 $ & $2.71$ $\sigma$ & $0.54$ $\sigma$  & $ \beta_{max}              = 0.79 $ & $2.80$ $\sigma$ & $1.00$ $\sigma$  \\ 
   \bottomrule
  \end{tabular}
	  \caption{The table summarizes the significance values of the observed periods for the local and global method based on the 
      monthly and weekly $\gamma$-ray light curves of PG 1553+113 and PKS 0447-489, respectively. 
      The significance levels are evaluated via MC-Simulation using Timmer \& Koenig 1995 and Emmanoulopoulos et al 2013 algorithms
      with 50000 trials. 
      The mean slopes $\langle \beta_{m} \rangle$ are obtained from fitting the $\gamma$-ray power spectra with simple power-law model, and the corresponding success 
     fraction are calculated based on the PSRESP method \citep{2002MNRAS.332..231U}. $\beta_{min}$ and $ \beta_{min}$ represents the minimum and maximum of the $1\sigma$ error bars 
     resulting from fitting Gaussians to each of the slope profile.
     }
      \label{table: significance dependence}
\end{table*}
\section{Discussion and Conclusions}
There are an increasing number of reports suggesting the presence of year-type periodicities in the $\gamma$-ray 
light curves of Fermi-LAT detected blazars. Given the still limited duration of the light curves ($\simlt 10$ yr), care 
needs to be taken, however, to properly assess the significance of these periods against a colored-noise background. 
\citet{2016MNRAS.461.3145V} have shown, for example, that clear phantom periodicities can be found even in pure noise 
data, with typical periods corresponding to $(1.5-2.5)$ cycles over the available data. 

Our systematic investigation performed in this study reveals that four out of the six investigated blazars show long-term QPO indications (local significance $\simgt 3\sigma$) in their Fermi-LAT light curves with an intrinsic period around $1.6$ yr. As we have shown there is some uncertainty as to the significance of these periodicities. Depending on the inferred best-fit PSD slope, the (local) significance can encompass a range from $\sim2.6\sigma$ to $>5\sigma$ (method 1), cf. Table.~\ref{table: significance}. The QPO significance is however strongly reduced in terms of a global significance, suggesting that longterm period claims should be treated with caution.

While all sources, except PG 1553+113, show clear indications for log-normality in their distribution of fluxes, 
incorporation of the appropriate simulation method (Timmer \& Koenig 1995 or Emmanoulopoulos et al. 2013) does not have a strong impact on the significance evaluation. Improving the PSD characterization, i.e. by narrowing down the range for the PSD slope will instead be more relevant to better assess the significance. 

Though the inferred periods are only tentative, one could speculate about physical mechanisms capable of 
accounting for year-type periodicity. It seems in fact surprising that the intrinsic periods appear quite similar, 
clustering around $P\sim 1.6$ yr for sources at different redshifts.

Perhaps one of the most natural scenarios for the origin of longterm periodic variability is related to the orbital 
motion in a supermassive binary black hole system (SBBHS). In principle a SBBHS phase is likely to occur in radio-loud 
AGN (being hosted by elliptically galaxies) at some stage. However, the periods inferred here appear rather short to 
plausibly relate them to orbital SBBH motion. Gravitation radiation would lead to coalescence on a characteristic timescale 
$t_{grav}\simeq 7 \times 10^3 \,(1+q)^{1/3}/[q \,(M_{BH}/5\times 10^8 M_{\odot})^{5/3}]\, (P/2~\mathrm{yr})^{8/3}$ 
yr only, so that for typical systems with $q:=m/M \geq 0.05$ the presumed source state would be highly unlikely. 
In fact, one rather expects orbital periods of the order $\sim 10$ yr for SBBHSs in blazars \citep{2007Ap&SS.309..271R}. 
This would then also be compatible with pulsar timing constraints on the inferred gravitational background \citep{2018MNRAS.481L..74H}. 
As orbital motion usually allows for the shortest periodic driving \citep[e.g.,][]{2004ApJ...615L...5R}, a direct SBBH 
origin of the observed periods appears less likely. This does not argue against the presence of SBBHSs in blazars in 
general, but simply cautions to directly relate periodicities of the
order of $P\sim 1$ yr to such systems.

Alternatively, QPOs might be related to quasi-periodic changes in accretion flow conditions that are effectively 
transmitted to the jet, modulating its non-thermal emission properties. Time-dependent modulations of the transition 
radius $r_{\rm t}$ between an outer cooling-dominated (standard) disc and an inner radiatively inefficient flow (ADAF), 
for example, could lead to periodic mass flux variations \cite[e.g.,][]{2003MNRAS.344..468G}. If one requires the
advective timescale $t_{\rm ad}(r_{\rm t}) \sim r_g\, (0.5\alpha\,c)^{-1} (r_{\rm t}/r_g)^{3/2}$, with $\alpha=0.25$ 
the viscosity coefficient and $r_g=GM_{\rm BH}/c^2$ the gravitational radius, to be (at most) comparable to $P$, this 
would place the transition radius at a characteristic scale of $r_{\rm t} \lesssim 200~r_g (P/1.6~\mathrm{yr})^{2/3}\,\
(5 \times 10^8 M_{\odot}/M_{\rm BH})^{2/3}$ for a reference mass of $M_{\rm BH}=5 \times 10^8 M_{\odot}$. This seems 
compatible with estimates for the transition radius in BL Lacs \citep[e.g.,][]{2003ApJ...599..147C,2008PASP..120..477X}. This would then suggest a similar black hole mass-scale for the systems investigated here.

On the other hand, year-type QPO could perhaps also trace plasma motion in the jet close to its outer jet radius $r_0$. In the lighthouse model \citep{1992A&A...255...59C}, for example, the disk-related jet is initially rotating, leading to a helical trajectory 
for a component injected on scales $r_0\sim 10\,r_L$ beyond the light cylinder $r_L\sim 10\,r_g$ of the innermost part of the disk magnetosphere. Angular momentum conservation would imply a characteristic intrinsic period for such a component of $P=2\pi r_L (r_0/r_L)^2/c \lesssim 2 \;(r_0/20r_L)^2 (M_{\rm BH}/5\times 10^8 M_{\odot})$ yr. While collimation might occur earlier \citep[e.g.,][]{1997A&A...319.1025F}, thereby reducing the jet radius, sligthly changing footpoint radii could possibly compensate for this. The lighthouse model was originally designed to account for observed QPOs with $P_{obs}\leq$ few weeks by the taking travel time effects with respect to an outwardly moving, (single) flaring component into account. It seems likely however, that the fundamental period $P$ might be visible even if the flux would be suppressed quickly. It will be interesting to probe this with a larger source sample, as intrinsic periods in this case are expected to be less than $2$ yr for a typcial black hole mass range. 

The fact that the intrinsic periods seems to be clustering around $P\sim 1.6$ yr remains particularly interesting and suggestive of a common physical mechanism. The inferred periods are, however, only tentative. Adding one or two more cycle of data (i.e., $\sim2-3$ yr) is expected to significantly improve the situation and to help to clarify their putative presence and physical implication.

\begin{acknowledgements}\\
    We would like to thank Stefan Wagner from Landessternwarte Heidelberg  (LSW) for comments and suggestions. FMR acknowledges support by a DFG
    Heisenberg Fellowship RI 1187/6-1.
\end{acknowledgements}

\bibliography{aa.bib}

\begin{thebibliography}{59}
\expandafter\ifx\csname natexlab\endcsname\relax\def\natexlab#1{#1}\fi

\bibitem[{{Abdo} {et~al.}(2010){Abdo}, {Ackermann}, {Ajello}, {Atwood},
  {Axelsson}, {Baldini}, {Ballet}, {Barbiellini}, {Baring}, \&
  {Bastieri}}]{2010ApJS..187..460A}
{Abdo}, A.~A., {Ackermann}, M., {Ajello}, M., {et~al.} 2010, \apjs, 187, 460

\bibitem[{{Acero} {et~al.}(2015){Acero}, {Ackermann}, {Ajello}, {Albert},
  {Atwood}, {Axelsson}, {Baldini}, {Ballet}, {Barbiellini}, {Bastieri},
  {Belfiore}, {Bellazzini}, {Bissaldi}, {Blandford}, {Bloom}, {Bogart},
  {Bonino}, {Bottacini}, {Bregeon}, {Britto}, {Bruel}, {Buehler}, {Burnett},
  {Buson}, {Caliandro}, {Cameron}, {Caputo}, {Caragiulo}, {Caraveo},
  {Casandjian}, {Cavazzuti}, {Charles}, {Chaves}, {Chekhtman}, {Cheung},
  {Chiang}, {Chiaro}, {Ciprini}, {Claus}, {Cohen-Tanugi}, {Cominsky}, {Conrad},
  {Cutini}, {D'Ammando}, {de Angelis}, {DeKlotz}, {de Palma}, {Desiante},
  {Digel}, {Di Venere}, {Drell}, {Dubois}, {Dumora}, {Favuzzi}, {Fegan},
  {Ferrara}, {Finke}, {Franckowiak}, {Fukazawa}, {Funk}, {Fusco}, {Gargano},
  {Gasparrini}, {Giebels}, {Giglietto}, {Giommi}, {Giordano}, {Giroletti},
  {Glanzman}, {Godfrey}, {Grenier}, {Grondin}, {Grove}, {Guillemot}, {Guiriec},
  {Hadasch}, {Harding}, {Hays}, {Hewitt}, {Hill}, {Horan}, {Iafrate}, {Jogler},
  {J{\'o}hannesson}, {Johnson}, {Johnson}, {Johnson}, {Johnson}, {Kamae},
  {Kataoka}, {Katsuta}, {Kuss}, {La Mura}, {Landriu}, {Larsson}, {Latronico},
  {Lemoine-Goumard}, {Li}, {Li}, {Longo}, {Loparco}, {Lott}, {Lovellette},
  {Lubrano}, {Madejski}, {Massaro}, {Mayer}, {Mazziotta}, {McEnery},
  {Michelson}, {Mirabal}, {Mizuno}, {Moiseev}, {Mongelli}, {Monzani},
  {Morselli}, {Moskalenko}, {Murgia}, {Nuss}, {Ohno}, {Ohsugi}, {Omodei},
  {Orienti}, {Orlando}, {Ormes}, {Paneque}, {Panetta}, {Perkins},
  {Pesce-Rollins}, {Piron}, {Pivato}, {Porter}, {Racusin}, {Rando}, {Razzano},
  {Razzaque}, {Reimer}, {Reimer}, {Reposeur}, {Rochester}, {Romani},
  {Salvetti}, {S{\'a}nchez-Conde}, {Saz Parkinson}, {Schulz}, {Siskind},
  {Smith}, {Spada}, {Spandre}, {Spinelli}, {Stephens}, {Strong}, {Suson},
  {Takahashi}, {Takahashi}, {Tanaka}, {Thayer}, {Thayer}, {Thompson},
  {Tibaldo}, {Tibolla}, {Torres}, {Torresi}, {Tosti}, {Troja}, {Van Klaveren},
  {Vianello}, {Winer}, {Wood}, {Wood}, {Zimmer}, \& {Fermi-LAT
  Collaboration}}]{2015ApJS..218...23A}
{Acero}, F., {Ackermann}, M., {Ajello}, M., {et~al.} 2015, \apjs, 218, 23

\bibitem[{{Ackermann} {et~al.}(2015){Ackermann}, {Ajello}, {Atwood}, {Baldini},
  {Ballet}, {Barbiellini}, {Bastieri}, {Becerra Gonzalez}, {Bellazzini},
  {Bissaldi}, {Blandford}, {Bloom}, {Bonino}, {Bottacini}, {Brandt}, {Bregeon},
  {Britto}, {Bruel}, {Buehler}, {Buson}, {Caliandro}, {Cameron}, {Caragiulo},
  {Caraveo}, {Carpenter}, {Casandjian}, {Cavazzuti}, {Cecchi}, {Charles},
  {Chekhtman}, {Cheung}, {Chiang}, {Chiaro}, {Ciprini}, {Claus},
  {Cohen-Tanugi}, {Cominsky}, {Conrad}, {Cutini}, {D'Abrusco}, {D'Ammando}, {de
  Angelis}, {Desiante}, {Digel}, {Di Venere}, {Drell}, {Favuzzi}, {Fegan},
  {Ferrara}, {Finke}, {Focke}, {Franckowiak}, {Fuhrmann}, {Fukazawa},
  {Furniss}, {Fusco}, {Gargano}, {Gasparrini}, {Giglietto}, {Giommi},
  {Giordano}, {Giroletti}, {Glanzman}, {Godfrey}, {Grenier}, {Grove},
  {Guiriec}, {Hewitt}, {Hill}, {Horan}, {Itoh}, {J{\'o}hannesson}, {Johnson},
  {Johnson}, {Kataoka}, {Kawano}, {Krauss}, {Kuss}, {La Mura}, {Larsson},
  {Latronico}, {Leto}, {Li}, {Li}, {Longo}, {Loparco}, {Lott}, {Lovellette},
  {Lubrano}, {Madejski}, {Mayer}, {Mazziotta}, {McEnery}, {Michelson},
  {Mizuno}, {Moiseev}, {Monzani}, {Morselli}, {Moskalenko}, {Murgia}, {Nuss},
  {Ohno}, {Ohsugi}, {Ojha}, {Omodei}, {Orienti}, {Orlando}, {Paggi}, {Paneque},
  {Perkins}, {Pesce-Rollins}, {Piron}, {Pivato}, {Porter}, {Rain{\`o}},
  {Rando}, {Razzano}, {Razzaque}, {Reimer}, {Reimer}, {Romani}, {Salvetti},
  {Schaal}, {Schinzel}, {Schulz}, {Sgr{\`o}}, {Siskind}, {Sokolovsky}, {Spada},
  {Spandre}, {Spinelli}, {Stawarz}, {Suson}, {Takahashi}, {Takahashi},
  {Tanaka}, {Thayer}, {Thayer}, {Tibaldo}, {Torres}, {Torresi}, {Tosti},
  {Troja}, {Uchiyama}, {Vianello}, {Winer}, {Wood}, \&
  {Zimmer}}]{2015ApJ...810...14A}
{Ackermann}, M., {Ajello}, M., {Atwood}, W.~B., {et~al.} 2015, \apj, 810, 14

\bibitem[{Ackermann {et~al.}(2015)}]{Ackermann:2015wda}
Ackermann, M. {et~al.} 2015, \apj, 813, L41

\bibitem[{{Aharonian} {et~al.}(2007){Aharonian}, {Akhperjanian},
  {et~al.}}]{2007ApJ...664L..71A}
{Aharonian}, F., {Akhperjanian}, {et~al.} 2007, \apjl, 664, L71

\bibitem[{Anderson \& Darling(1954)}]{ad1954}
Anderson, T.~W. \& Darling, D.~A. 1954, Journal of the American Statistical
  Association, 49, 765

\bibitem[{{Ar{\'e}valo} \& {Uttley}(2006)}]{2006MNRAS.367..801A}
{Ar{\'e}valo}, P. \& {Uttley}, P. 2006, \mnras, 367, 801

\bibitem[{{Atwood} {et~al.}(2013)}]{2013arXiv1303.3514A}
{Atwood}, W. {et~al.} 2013, ArXiv e-prints [\eprint[arXiv]{1303.3514}]

\bibitem[{{Bell} {et~al.}(2011){Bell}, {Tzioumis}, {Uttley}, {Fender},
  {Ar{\'e}valo}, {Breedt}, {McHardy}, {Calvelo}, {Jamil}, \&
  {K{\"o}rding}}]{2011MNRAS.411..402B}
{Bell}, M.~E., {Tzioumis}, T., {Uttley}, P., {et~al.} 2011, \mnras, 411, 402

\bibitem[{{Bhatta} {et~al.}(2016){Bhatta}, {Zola}, {Stawarz}, {Ostrowski},
  {Winiarski}, {Og{\l}oza}, {Dr{\'o}{\.z}d{\.z}}, {Siwak}, {Liakos},
  {Kozie{\l}-Wierzbowska}, {Gazeas}, {Debski}, {Kundera}, {Stachowski}, \&
  {Paliya}}]{2016ApJ...832...47B}
{Bhatta}, G., {Zola}, S., {Stawarz}, {\L}., {et~al.} 2016, \apj, 832, 47

\bibitem[{{B{\"o}ttcher} \& {Chiang}(2002)}]{2002ApJ...581..127B}
{B{\"o}ttcher}, M. \& {Chiang}, J. 2002, \apj, 581, 127

\bibitem[{{Bravo} {et~al.}(2014){Bravo}, {Roque}, {Estrela}, {Le{\~a}o}, \& {De
  Medeiros}}]{2014A&A...568A..34B}
{Bravo}, J.~P., {Roque}, S., {Estrela}, R., {Le{\~a}o}, I.~C., \& {De
  Medeiros}, J.~R. 2014, \aap, 568, A34

\bibitem[{{Camenzind} \& {Krockenberger}(1992)}]{1992A&A...255...59C}
{Camenzind}, M. \& {Krockenberger}, M. 1992, \aap, 255, 59

\bibitem[{{Cao}(2003)}]{2003ApJ...599..147C}
{Cao}, X. 2003, \apj, 599, 147

\bibitem[{{Chatterjee} {et~al.}(2008){Chatterjee}, {Jorstad}, {Marscher}, {Oh},
  {McHardy}, {Aller}, {Aller}, {Balonek}, {Miller}, {Ryle}, {Tosti},
  {Kurtanidze}, {Nikolashvili}, {Larionov}, \&
  {Hagen-Thorn}}]{2008ApJ...689...79C}
{Chatterjee}, R., {Jorstad}, S.~G., {Marscher}, A.~P., {et~al.} 2008, \apj,
  689, 79

\bibitem[{{Covino} {et~al.}(2019){Covino}, {Sandrinelli}, \&
  {Treves}}]{2019MNRAS.482.1270C}
{Covino}, S., {Sandrinelli}, A., \& {Treves}, A. 2019, \mnras, 482, 1270

\bibitem[{D'Agostino \& Pearson(1973)}]{doi:10.1093/biomet/60.3.613}
D'Agostino, R. \& Pearson, E.~S. 1973, Biometrika, 60, 613

\bibitem[{D'Agostino(1970)}]{doi:10.1093/biomet/57.3.679}
D'Agostino, R.~B. 1970, Biometrika, 57, 679

\bibitem[{{Edelson} {et~al.}(2002){Edelson}, {Turner}, {Pounds}, {Vaughan},
  {Markowitz}, {Marshall}, {Dobbie}, \& {Warwick}}]{2002ApJ...568..610E}
{Edelson}, R., {Turner}, T.~J., {Pounds}, K., {et~al.} 2002, \apj, 568, 610

\bibitem[{{Emmanoulopoulos} {et~al.}(2013){Emmanoulopoulos}, {McHardy}, \&
  {Papadakis}}]{2013MNRAS.433..907E}
{Emmanoulopoulos}, D., {McHardy}, I.~M., \& {Papadakis}, I.~E. 2013, \mnras,
  433, 907

\bibitem[{{Fan} {et~al.}(2002){Fan}, {Lin}, {Xie}, {Zhang}, {Mei}, {Su}, \&
  {Peng}}]{2002A&A...381....1F}
{Fan}, J.~H., {Lin}, R.~G., {Xie}, G.~Z., {et~al.} 2002, \aap, 381, 1

\bibitem[{{Fendt}(1997)}]{1997A&A...319.1025F}
{Fendt}, C. 1997, \aap, 319, 1025

\bibitem[{{Foster}(1996)}]{1996AJ....112.1709F}
{Foster}, G. 1996, \aj, 112, 1709

\bibitem[{{Gracia} {et~al.}(2003){Gracia}, {Peitz}, {Keller}, \&
  {Camenzind}}]{2003MNRAS.344..468G}
{Gracia}, J., {Peitz}, J., {Keller}, C., \& {Camenzind}, M. 2003, \mnras, 344,
  468

\bibitem[{{Gross} \& {Vitells}(2010)}]{2010EPJC...70..525G}
{Gross}, E. \& {Vitells}, O. 2010, European Physical Journal C, 70, 525

\bibitem[{Han \& van~der Baan(2013)}]{doi:10.1190/geo2012-0199.1}
Han, J. \& van~der Baan, M. 2013, GEOPHYSICS, 78, O9

\bibitem[{{Holgado} {et~al.}(2018){Holgado}, {Sesana}, {Sandrinelli}, {Covino},
  {Treves}, {Liu}, \& {Ricker}}]{2018MNRAS.481L..74H}
{Holgado}, A.~M., {Sesana}, A., {Sandrinelli}, A., {et~al.} 2018, \mnras, 481,
  L74

\bibitem[{{Kadler} {et~al.}(2006){Kadler}, {Hughes}, {Ros}, {Aller}, \&
  {Aller}}]{2006A&A...456L...1K}
{Kadler}, M., {Hughes}, P.~A., {Ros}, E., {Aller}, M.~F., \& {Aller}, H.~D.
  2006, \aap, 456, L1

\bibitem[{{King} {et~al.}(2013){King}, {Hovatta}, {Max-Moerbeck}, {Meier},
  {Pearson}, {Readhead}, {Reeves}, {Richards}, \&
  {Shepherd}}]{2013MNRAS.436L.114K}
{King}, O.~G., {Hovatta}, T., {Max-Moerbeck}, W., {et~al.} 2013, \mnras, 436,
  L114

\bibitem[{Li {et~al.}(2009)Li, Xie, Chen, Dai, Lei, Yi, \&
  Ren}]{1538-3873-121-885-1172}
Li, H.~Z., Xie, G.~Z., Chen, L.~E., {et~al.} 2009, \pasp, 121, 1172

\bibitem[{{Lomb}(1976)}]{1976Ap&SS..39..447L}
{Lomb}, N.~R. 1976, \apss, 39, 447

\bibitem[{Lyons(2008)}]{Lyons:1900zz}
Lyons, L. 2008, Ann. Appl. Stat., 2, 887

\bibitem[{{Lyubarskii}(1997)}]{1997MNRAS.292..679L}
{Lyubarskii}, Y.~E. 1997, \mnras, 292, 679

\bibitem[{{Mattox} {et~al.}(1996){Mattox}, {Bertsch}, {Chiang}, {Dingus},
  {Digel}, {Esposito}, {Fierro}, {Hartman}, {Hunter}, {Kanbach}, {Kniffen},
  {Lin}, {Macomb}, {Mayer-Hasselwander}, {Michelson}, {von Montigny},
  {Mukherjee}, {Nolan}, {Ramanamurthy}, {Schneid}, {Sreekumar}, {Thompson}, \&
  {Willis}}]{1996ApJ...461..396M}
{Mattox}, J.~R., {Bertsch}, D.~L., {Chiang}, J., {et~al.} 1996, \apj, 461, 396

\bibitem[{{Mohan} \& {Mangalam}(2015)}]{2015ApJ...805...91M}
{Mohan}, P. \& {Mangalam}, A. 2015, \apj, 805, 91

\bibitem[{Prokhorov \& Moraghan(2017)}]{Prokhorov:2017amk}
Prokhorov, D.~A. \& Moraghan, A. 2017, \mnras, 471, 3036

\bibitem[{{Rieger}(2004)}]{2004ApJ...615L...5R}
{Rieger}, F.~M. 2004, \apjl, 615, L5

\bibitem[{{Rieger}(2007)}]{2007Ap&SS.309..271R}
{Rieger}, F.~M. 2007, \apss, 309, 271

\bibitem[{Sandrinelli {et~al.}(2014)Sandrinelli, Covino, \&
  Treves}]{2041-8205-793-1-L1}
Sandrinelli, A., Covino, S., \& Treves, A. 2014, \apjl, 793, L1

\bibitem[{Sandrinelli {et~al.}(2016)Sandrinelli, Covino, \&
  Treves}]{Sandrinelli:2015ijk}
Sandrinelli, A., Covino, S., \& Treves, A. 2016, \apj, 820, 20

\bibitem[{{Scargle}(1982)}]{1982ApJ...263..835S}
{Scargle}, J.~D. 1982, \apj, 263, 835

\bibitem[{{Shah} {et~al.}(2018){Shah}, {Mankuzhiyil}, {Sinha}, {Misra},
  {Sahayanathan}, \& {Iqbal}}]{Shah2018}
{Shah}, Z., {Mankuzhiyil}, N., {Sinha}, A., {et~al.} 2018, Research in
  Astronomy and Astrophysics, 18, 141

\bibitem[{Shapiro \& Wilk(1965)}]{10.2307/2333709}
Shapiro, S.~S. \& Wilk, M.~B. 1965, Biometrika, 52, 591

\bibitem[{{Sobacchi} {et~al.}(2017){Sobacchi}, {Sormani}, \&
  {Stamerra}}]{2017MNRAS.465..161S}
{Sobacchi}, E., {Sormani}, M.~C., \& {Stamerra}, A. 2017, \mnras, 465, 161

\bibitem[{{Stephens}(1974)}]{stephens1976}
{Stephens}, M.~A. 1974, Journal of the American Statistical Association, 69,
  730

\bibitem[{{Timmer} \& {Koenig}(1995)}]{1995A&A...300..707T}
{Timmer}, J. \& {Koenig}, M. 1995, \aap, 300, 707

\bibitem[{{Torrence} \& {Compo}(1998)}]{1998BAMS...79...61T}
{Torrence}, C. \& {Compo}, G.~P. 1998, Bulletin of the American Meteorological
  Society, 79, 61

\bibitem[{{Urry} \& {Padovani}(1995)}]{1995PASP..107..803U}
{Urry}, C.~M. \& {Padovani}, P. 1995, \pasp, 107, 803

\bibitem[{{Uttley} {et~al.}(2002){Uttley}, {McHardy}, \&
  {Papadakis}}]{2002MNRAS.332..231U}
{Uttley}, P., {McHardy}, I.~M., \& {Papadakis}, I.~E. 2002, \mnras, 332, 231

\bibitem[{{Uttley} {et~al.}(2005){Uttley}, {McHardy}, \&
  {Vaughan}}]{2005MNRAS.359..345U}
{Uttley}, P., {McHardy}, I.~M., \& {Vaughan}, S. 2005, \mnras, 359, 345

\bibitem[{{Valtonen} {et~al.}(2006){Valtonen}, {Lehto}, {Sillanp{\"a}{\"a}},
  {Nilsson}, {Mikkola}, {Hudec}, {Basta}, {Ter{\"a}sranta}, {Haque}, \&
  {Rampadarath}}]{2006ApJ...646...36V}
{Valtonen}, M.~J., {Lehto}, H.~J., {Sillanp{\"a}{\"a}}, A., {et~al.} 2006,
  \apj, 646, 36

\bibitem[{{Vaughan}(2010)}]{2010MNRAS.402..307V}
{Vaughan}, S. 2010, \mnras, 402, 307

\bibitem[{{Vaughan} {et~al.}(2003){Vaughan}, {Edelson}, {Warwick}, \&
  {Uttley}}]{2003MNRAS.345.1271V}
{Vaughan}, S., {Edelson}, R., {Warwick}, R.~S., \& {Uttley}, P. 2003, \mnras,
  345, 1271

\bibitem[{{Vaughan} {et~al.}(2016){Vaughan}, {Uttley}, {Markowitz},
  {Huppenkothen}, {Middleton}, {Alston}, {Scargle}, \&
  {Farr}}]{2016MNRAS.461.3145V}
{Vaughan}, S., {Uttley}, P., {Markowitz}, A.~G., {et~al.} 2016, \mnras, 461,
  3145

\bibitem[{{Wiita}(2011)}]{2011JApA...32..147W}
{Wiita}, P.~J. 2011, Journal of Astrophysics and Astronomy, 32, 147

\bibitem[{{Xie} {et~al.}(2008){Xie}, {Hao}, {Du}, {Zhang}, \&
  {Jia}}]{2008PASP..120..477X}
{Xie}, Z.~H., {Hao}, J.~M., {Du}, L.~M., {Zhang}, X., \& {Jia}, Z.~L. 2008,
  \pasp, 120, 477

\bibitem[{{Zechmeister} \& {K{\"u}rster}(2009)}]{2009A&A...496..577Z}
{Zechmeister}, M. \& {K{\"u}rster}, M. 2009, \aap, 496, 577

\bibitem[{Zhang {et~al.}(2017)Zhang, Yan, Liao, Zeng, Wang, \&
  Cao}]{Zhang:2017ear}
Zhang, P., Yan, D., Liao, N., {et~al.} 2017, {\apj}, 842, 10

\bibitem[{{Zhang} {et~al.}(2017){Zhang}, {Yan}, {Zhou}, {Fan}, {Wang}, \&
  {Zhang}}]{2017ApJ...845...82Z}
{Zhang}, P.-F., {Yan}, D.-H., {Zhou}, J.-N., {et~al.} 2017, \apj, 845, 82

\end{thebibliography}

\end{document}